\documentclass[11pt,a4paper]{article}
\pdfoutput=1                   % to allow pdflatex compilation in JCAP
\usepackage{jheppub}
\usepackage{mathtools}         % Fixes and extends (and also loads) amsmath.
\usepackage{graphicx}
\usepackage{mathtools}        % Fixes and extends (and also loads) amsmath.
\usepackage{bbm}              % Blackboard math variants of CM fonts. (For \mathbbm{1}.)
\graphicspath{{./Figs/}}
\usepackage{comment}
\usepackage{latexsym}
\usepackage{xspace}
\usepackage{lscape}
\usepackage{color}
\usepackage{hyperref}
\usepackage{bm}
\usepackage{bbm}
\usepackage{relsize}
\usepackage{tabularx}
\usepackage{multirow}
\usepackage{amsmath}
\usepackage{amssymb}
\usepackage[table]{xcolor}
\usepackage{lipsum}
\usepackage{feynmp-auto}
\usepackage{upgreek}
\usepackage{nicefrac}
\usepackage[titletoc]{appendix}

\usepackage{caption}
\usepackage{subcaption}

\usepackage{cleveref} % better cross-refs using \cref (load after hyperref!)
    \crefname{equation}{}{}
    \crefname{align}{}{}
    \crefname{figure}{}{}    % example: \crefformat{figure}{#figure~#1#3}
    \crefname{table}{}{}
    \crefname{section}{}{}   % example: \crefformat{section}{#2section~#1#3}
    \crefname{appendix}{}{}
    \crefname{footnote}{}{}
    
    \crefalias{subequation}{equation}

\newcommand{\sfrac}[2]{{\textstyle{\frac{#1}{#2}}}} % small fraction (for display math)
\newcommand{\overbar}[1]{\mkern 1.5mu\overline{\mkern-1.5mu#1\mkern-1.5mu}\mkern 1.5mu} % finer-sized bar/overline

\def\beq{\begin{equation}}
\def\eeq{\end{equation}}
\def\baq{\begin{eqnarray}}
\def\eaq{\end{eqnarray}}

\DeclareMathOperator{\dalembert}{\square}

\title{Tachyonic particle production: quantum 2PI formalism with momentum exchanging collisions\\
}

\author[a,b]{Kimmo Kainulainen}
\author[a,b]{Sami Nurmi}
\author[a,b]{Olli Väisänen}

\affiliation[a]{Department of Physics, PL 35 (YFL), 40014 University of Jyv\"askyl\"a, Finland}
\affiliation[b]{Helsinki Institute of Physics, PL 64, 00014 University of Helsinki, Finland}
\emailAdd{kimmo.kainulainen@jyu.fi}
\emailAdd{olli.j.r.vaisanen@jyu.fi}
\emailAdd{sami.t.nurmi@jyu.fi}

\abstract{Oscillating spacetime curvature can drive particle production during reheating, whose accurate modeling requires the use of non-perturbative out-of-equilibrium methods. Tachyonic instabilities have previously been studied using 2-Particle Irreducible (2PI) formalism in the Hartree approximation, which however misses important momentum exchanging interactions. We present a self-consistent approximation scheme for reducing the non-local next-to-leading order 2PI equations of motion to local quantum kinetic equations, which can be solved with standard methods. We pay special attention to interactions involving unstable modes during tachyonic instabilities.}


\begin{document}

%\arxivnumber{XXXX.XXXXX}
\maketitle

%%%%%%%%%%%%%%%%%%%%%%%%%%%%%%%%%%%%%%%%%%%%%%%%%%%%%%%%%%%%%%%%%%%%%%%%%%%%%%%%%%%%%%%%%%%%%%%%%%%%%
%%%%%%%%%%%%%%%%%%%%%%%%%%%%%%%%%%%%%%%%%%%%%%%%%%%%%%%%%%%%%%%%%%%%%%%%%%%%%%%%%%%%%%%%%%%%%%%%%%%%%
%
\section{Introduction}
\label{sec:intro}
%
%%%%%%%%%%%%%%%%%%%%%%%%%%%%%%%%%%%%%%%%%%%%%%%%%%%%%%%%%%%%%%%%%%%%%%%%%%%%%%%%%%%%%%%%%%%%%%%%%%%%%
%%%%%%%%%%%%%%%%%%%%%%%%%%%%%%%%%%%%%%%%%%%%%%%%%%%%%%%%%%%%%%%%%%%%%%%%%%%%%%%%%%%%%%%%%%%%%%%%%%%%%

Many important processes in the early universe involve out-of-equilibrium quantum dynamics of coherent scalar fields in the presence of decohering interactions. Some key examples include resonant   phenomena~\cite{Kofman:1994rk,kofman:1997yn,Greene:1997fu,Braden:2010wd,Berges:2002cz} and tachyonic instabilities~\cite{Calzetta:1989bj,Guth:1985ya,Weinberg:1987vp,Bassett:1997az,Felder:2000hj,Felder:2001kt,Dufaux:2006ee} in reheating, various non-thermal mechanisms for dark matter production ~\cite{Kolb:1998ki,Chung:1998zb,Garny:2015sjg,Garny:2017kha,Tang:2016vch,Markkanen:2015xuw, Fairbairn:2018bsw, Cembranos:2019qlm}, the electroweak baryogenesis~\cite{Cline:2000nw,Kainulainen:2001cn,Kainulainen:2002th,Cline:2013gha,Cline:2020jre,Konstandin:2013caa,Kainulainen:2021oqs} and the leptogenesis mechanism~\cite{Buchmuller:2000nd,Beneke:2010dz,Anisimov:2010dk,Dev:2017trv,DeSimone:2007gkc,Garny:2009qn,Garbrecht:2011aw,Garny:2011hg,Dev:2017wwc,Jukkala:2021sku}. Non-perturbative methods, such as the Schwinger-Keldysh 2PI approach~\cite{Cornwall:1974vz,Berges:2004yj}, are often needed for comprehensive modelling of non-linear quantum dynamics in such setups ~\cite{Boyanovsky:1992vi,Boyanovsky:1993pf,Baacke:2001zt,Berges:2002wr,Arrizabalaga:2004iw,Arrizabalaga:2005tf,Kainulainen:2021eki,Tranberg:2023uzs}.  

In~\cite{Kainulainen_2023, Kainulainen:2024etd} the 2PI methods in the lowest order Hartree truncation were applied for the dark matter scenario of \cite{Markkanen:2015xuw,Fairbairn:2018bsw}, where a scalar singlet with a non-minimal coupling $\xi R \chi^2$ undergoes tachyonic instabilities during reheating, when the curvature scalar $R$  oscillates between positive and negative values. The excitations produced via this gravitational coupling can constitute the observed dark matter relic abundance even if the singlet is completely decoupled from the visible matter \cite{Markkanen:2015xuw,Fairbairn:2018bsw}. The 2PI Hartee results of \cite{Kainulainen_2023} showed that scalar self-interactions $\lambda \chi^{4}$ can trigger strong resonant instabilities at the end of the tachyonic stage, and qualitatively similar effects were seen in the classical lattice simulations performed in \cite{Kainulainen:2024etd}. The Hartree approximation however does not account for momentum exchanging processes in the quantum transport equations which describe the evolution of the system. Such processes were found to be dynamically relevant in the classical lattice evolution in \cite{Kainulainen:2024etd}, but it is not a priori clear how well the classical approximation describes the phase space evolution in this dark matter setup, which features non-linear out-of-equilibrium evolution of coherent quantum states. 

In this work we develop the application of quantum 2PI methods for self-interacting scalar fields beyond the Hartree approximation, accounting for dynamical effects of to the collision terms. We use the dark matter scenario of~\cite{Markkanen:2015xuw,Fairbairn:2018bsw} as a concrete template for setting up the formalism but our methods are general and can be straightforwardly applied to any scalar model. Our approach is based on the moment expansion of the exact 2PI equations formulated for scalar fields in~\cite{Herranen:2008yg,Fidler:2011yq}. The coupled non-linear moment equations contain non-local collision integrals which in practice cannot be solved without further approximations. To render the problem into a computationally feasible form we approximate the two-point functions within the collision integrals by an Ansatz derived from leading order gradient expansion, similar to methods applied  in~\cite{Fidler:2011yq}. We formulate our Ansatz separately for stable modes with positive effective mass squared and for tachyonic modes with negative mass squared, corresponding to the tachyonic regimes in the dark matter scenario of~\cite{Markkanen:2015xuw,Fairbairn:2018bsw}.  With this approach we derive explicit expressions for the collision integrals, and present renormalized quantum transport equations which account for momentum exchanging collision processes and are formulated in terms of local moments the two-point function. Our equations can be solved numerically using methods similar to momentum-dependent Boltzmann equations. 

An alternative strategy would be to solve the 2PI equations directly on lattice~\cite{Berges:2002wr,Tranberg:2023uzs}, with no other approximations than truncating the infinite set of equations in some coupling constant expansion. While this approach is in principle more fundamental than ours, the lattice simulations are computationally expensive and their resolution is limited by numerically accessible simulation volumes. The lattice approach may therefore not be practical for setups like~\cite{Fairbairn:2018bsw}, where evolution of the system has to be followed over a long time, a wide range of momentum scales affects the dynamics, and details of the final phase space distribution are phenomenologically relevant. Indeed, even the classical lattice simulations in~\cite{Kainulainen:2024etd} were not able to fully track the final stages of the evolution in this setup and the computational challenges would be even more pronounced in quantum simulations. In contrast, our 2PI methods are substantially simpler to implement numerically and should be ideally suited for setups like~\cite{Fairbairn:2018bsw}. Concrete numerical work is beyond the scope of this work however, and we will return to it elsewhere~\cite{KNV4}.

This paper is organized as follows. In section~\cref{sec:model} we briefly specify the setup. In section~\cref{sec:2PI} we write down the general Kadanoff-Baym equations and specify the approximations under which they reduce to a local aquantum kinetic equation in the form of a coupled set of moment equations. In section~\cref{sec:ASA_approximation} we present the lowest order gradient expansion Ansatz, which renders the non-local collision integrals into a computationally feasible form.  In section~\cref{sec:collision_integral} we evaluate the collision integrals for stable modes and in section ~\cref{sec:Landau_damping} for stable tachyonic modes with real valued frequencies. In section ~\cref{sec:ASA_approximation_unstable} we evaluate the collision integrals for unstable tachyonic modes with imaginary valued frequencies. Finally, in section~\cref{sec:conclusions} we present our conclusions.

%%%%%%%%%%%%%%%%%%%%%%%%%%%%%%%%%%%%%%%%%%%%%%%%%%%%%%%%%%%%%%%%%%%%%%%%%%%%%%%%%%%%%%%%%%%%%%%%%%%%%
%%%%%%%%%%%%%%%%%%%%%%%%%%%%%%%%%%%%%%%%%%%%%%%%%%%%%%%%%%%%%%%%%%%%%%%%%%%%%%%%%%%%%%%%%%%%%%%%%%%%%
%
\section{The setup}
\label{sec:model}
%
%%%%%%%%%%%%%%%%%%%%%%%%%%%%%%%%%%%%%%%%%%%%%%%%%%%%%%%%%%%%%%%%%%%%%%%%%%%%%%%%%%%%%%%%%%%%%%%%%%%%%
%%%%%%%%%%%%%%%%%%%%%%%%%%%%%%%%%%%%%%%%%%%%%%%%%%%%%%%%%%%%%%%%%%%%%%%%%%%%%%%%%%%%%%%%%%%%%%%%%%%%%

While the methods developed in this work are generally applicable, we will use the dark matter setup of~\cite{Fairbairn:2018bsw} as a motivation and a concrete example. This setup consists of a non-minimally coupled spectator scalar singlet $\chi$ whose action is given by  
\begin{equation}
    \mathcal S_\chi = \int d^4 x \sqrt{-g} \left[\frac12 (\nabla^\mu \chi)(\nabla_\mu \chi) 
    - \frac12 m_\chi^2 \chi^2 + \frac{\xi}{2} R \chi^2 - \frac{\lambda}{4!} \chi^4 \right]~. 
\label{eq:action}
\end{equation}
We use the $(+,-,-,-)$ sign convention following~\cite{Kainulainen_2023}. The metric is given by the homogeneous and isotropic, spatially flat RW solution, and $R$ denotes the Ricci scalar. We recast the kinetic term of~\cref{eq:action} into a flat spacetime form by rescaling the field with the scale factor, $\chi = \sigma/a$, and switching to the conformal time. Somewhat unconventionally we will denote conformal time by $t$, defining $a \mathrm{d}t \equiv \mathrm{d} x^0$. This gives
\beq
\mathcal{S}_{\sigma} = \int \mathrm{d}t \, \mathrm{d}^3\bm{x}\left[\frac{1}{2}(\partial_t\sigma)^2-\frac{1}{2}(\nabla\sigma)^2
 -\frac{1}{2} m_{\rm eff}^2(t) \sigma^2-\frac{\lambda}{4!}\sigma^4\right],
\label{eq:sigma_action}
\eeq
where the effective mass term is
\begin{equation}
m^2_{\rm eff}(t) \equiv a^2(t)\bigg[m^2_{\chi} - \bigg( \xi-\frac{1}{6} \bigg)R(t)\bigg].
\label{eq:effective_mass}
\end{equation}
In the following, all correlation functions are computed with respect to the scaled field $\sigma$. 

In the specific setup of~\cite{Fairbairn:2018bsw}, the scale factor $a(t)$ and the Ricci scalar $R(t)$ are determined by dynamics of the inflaton field during reheating, and $R(t)$ oscillates from positive to negative values. This leads to tachyonic instability of the $\sigma$ field and generates a dark matter component, see~\cite{Fairbairn:2018bsw, Kainulainen_2023, Kainulainen:2024etd} for details. 

In this work, we treat~(\ref{eq:sigma_action}) simply as a general effective action that sets the starting point for our analysis. Our focus is in setting up a general theoretical framework for analysing the coupled evolution of one- and two-point correlators of the field $\sigma$ in the presence of momentum exchanging interactions. However, since we want our methods to be applicable also for the specific setup of~\cite{Fairbairn:2018bsw}, it is necessary to provide a formalism that is sufficiently general to handle cases where~\cref{eq:effective_mass} periodically takes negative values and $\sigma$ undergoes transient tachyonic instabilities. Apart from this consideration, the scale factor and the Ricci scalar in~\cref{eq:effective_mass} are just some external functions whose precise form is not relevant in our analysis. 

%%%%%%%%%%%%%%%%%%%%%%%%%%%%%%%%%%%%%%%%%%%%%%%%%%%%%%%%%%%%%%%%%%%%%%%%%%%%%%%%%%%%%%%%%%%%%%%%%%%%%
%%%%%%%%%%%%%%%%%%%%%%%%%%%%%%%%%%%%%%%%%%%%%%%%%%%%%%%%%%%%%%%%%%%%%%%%%%%%%%%%%%%%%%%%%%%%%%%%%%%%%
%
\section{The quantum 2PI-equations}
\label{sec:2PI}
%
%%%%%%%%%%%%%%%%%%%%%%%%%%%%%%%%%%%%%%%%%%%%%%%%%%%%%%%%%%%%%%%%%%%%%%%%%%%%%%%%%%%%%%%%%%%%%%%%%%%%%
%%%%%%%%%%%%%%%%%%%%%%%%%%%%%%%%%%%%%%%%%%%%%%%%%%%%%%%%%%%%%%%%%%%%%%%%%%%%%%%%%%%%%%%%%%%%%%%%%%%%%

We derive the equations of motion for the mean $\sigma$-field and its two-point function corresponding to the action~\cref{eq:sigma_action}, using the 2PI effective action methods~\cite{Cornwall:1974vz,Berges:2004yj}. For the single scalar field, the generic 2PI effective action reads as
\begin{equation}
\Gamma_{\rm 2PI} [\bar\sigma, \Delta] = \mathcal{S}[\bar\sigma] - \frac{i}{2}\mathrm{Tr}_{\mathcal{C}}\bigl[ \ln (\Delta)\bigr] + \frac{i}{2}\mathrm{Tr}_{\mathcal{C}}\bigl[\Delta_{0}^{-1}\Delta\bigr] 
+ \Gamma_2[\bar\sigma, \Delta].
\label{2PI-effective-action}
\end{equation}
Here $\mathcal{S}$ is the classical action and $\bar\sigma(x)$ and $i\Delta(x,y)$ are the classical one-point and the connected two-point functions of the scaled $\sigma$-field, defined over the complex Keldysh time contour $\mathcal{C}$~\cite{Keldysh:1964ud}. The traces in particular contain an integration over the complex time. The equations of motion follow from stationarity of the action~\cref{2PI-effective-action} with respect to variations in the field and the 2-point function. The real time equation of motion for the mean field $\bar\sigma$ in the example setup reads:

\begin{equation}
\Big[ \Box_x + m^2_{\rm eff}(x) + \frac{1}{6}\lambda\bar\sigma^2(x) 
+ \frac{1}{2} \lambda i\Delta(x,x) \Big] \bar\sigma(x) = \frac{\delta \Gamma_2}{\delta \bar\sigma(x)}.
\label{eq:eom-for-one-point-function}
\end{equation}
We left out the branch indices in the local correlation function $\Delta(x,x)$, because it is the same for all components of the two-point function $\Delta^{ab}(x,y)$. For the real-time two-point correlation function $\Delta^{ab}$, one finds the equation
\begin{equation}
\Delta_0^{-1}(x,y;\bar\sigma) \Delta^{ac}(x,y)
= a\delta^{ac} \delta^{(4)}(x-y) 
+ b\hspace{-.2em}\int \hspace{-.2em} \mathrm{d}^4z \, \Pi^{ab}(x,z)\Delta^{bc}(z,y),
\label{eq:eom-for-two-point-function}
\end{equation}
where summation over the Keldysh branch index $b=\pm$ is implied. The classical real-time tree-level inverse propagator for the example setup~\cref{eq:sigma_action} is defined as
\begin{equation}
\Delta_0^{-1}(x,y;\bar\sigma) = - \Big[ \dalembert_x + m^2_{\rm eff}(x) + \sfrac{1}{2}\lambda \bar\sigma^2\Big]\delta^{(4)}(x-y)
\label{eq:free-inverse-propagator}
\end{equation}
and the self-energy function is given by
\begin{equation}
  \Pi^{ab}(x,y) = 2\mathrm{i}ab \, \frac{\delta \Gamma_2[\bar\sigma, \Delta]}{\delta \Delta^{ba}(y,x)}.
\label{eq:general-self-energy}
\end{equation}
The equation~\cref{eq:eom-for-two-point-function} for the two-point function actually contains four equations. They are not all independent, however, and it is useful to decompose them into the following set of Kadanoff-Baym (KB) equations
\begin{align}
    (\Delta^{-1}_0 - \Pi^{r,a}) \otimes \Delta^{r,a} &= \delta, 
    \label{eq:pole}\\
    (\Delta^{-1}_0 - \Pi_{\rm H}) \otimes \Delta^<  &= \Pi^< \otimes \Delta_{\rm H} + \langle C \rangle, \label{eq:kadanoff_baym}
\end{align}
where $A\otimes B(x,y) \equiv \int {\rm d}^4z A(x,z)B(z,y)$. We suppressed the space-time indices for simplicity and defined the Wightman functions $\Delta^< = \Delta^{+-}$ and $\Delta^> = \Delta^{-+}$, as well as the retarded and advanced functions $\Delta^r = \Delta^{++}-\Delta^{<} =\Delta_{\rm F}-\Delta^<$ and $\Delta^a = \Delta_{\rm F}-\Delta^>$. The latter can also be broken into the Hermitian and anti-Hermitian parts: $\Delta^{r,a} = \Delta_{\rm H} \mp i{\cal A}$, where ${\cal A}$ is the spectral function defined as ${\cal A} = i(\Delta^> - \Delta^<)/2$. The same notations apply for relating the self-energy functions $\Pi^{ab}$ to $\Pi^{<,>}$ and $\Pi^{r,a} = \Pi_{\rm H} \mp i\Gamma$, where $\Gamma = i(\Pi^> - \Pi^<)/2$. The second term on the right hand-side of equation~\cref{eq:kadanoff_baym} is the collision integral
\begin{equation}
  \langle C \rangle = \frac12 \left( \Pi^> \otimes \Delta^< - \Pi^< \otimes \Delta^>  \right),
\label{eq:collision_integral}
\end{equation}
whose computation will make the bulk of this paper. 

\paragraph{Key approximations.}
Any practical solution of the coupled Kadanoff-Baym equations, (and the equation for the background field~\cref{eq:eom-for-one-point-function}), requires a set of approximations, which still should not reduce their ability to describe the relevant physics. Our scheme proceeds in the following steps:
\begin{enumerate}
    \item We decouple the (pole) equations for $\Delta^{r,a}$ from the (statistical) equation for $\Delta^<$.
    \item We use a different approximations for different self-energy functions. The Hermitian self-energy $\Pi_{\rm H}$ is computed at one loop (Hartree) level, while the width $\Gamma$ is neglected and the self-energies $\Pi^{<,>}$ are computed beyond Hartree to two loops.
    \item We use a moment expansion in Wigner space on the evolution equations.
    \item We employ the lowest order gradient expansion within the collision integrals.
\end{enumerate}

\noindent Decoupling the pole and statistical equations is essential to scheme to reduce the Kadanoff-Baym equations to local limit. We state only the main conceptual issue here and refer the interested reader to ref.~\cite{Kainulainen:2023ocv} for more details. The key observation is that the pole equations are inhomogeneous even in the collisionless limit and depend on the statistical function only indirectly through the self-energies $\Pi^{r,a}$. They can then be solved formally independently of the non-equilibrium statistical function, which only affects the values of $\Pi^{r,a}$, but not the {\rm form} of the pole solutions. Physically the pole functions describe the phase space structure of the theory rather than its kinematics. 

Next observe that while the decay rate $\Gamma$ and the self-energies $\Pi^{<,>}$ appear on same footing in KB-equations, their roles are completely different in decoupled equations. While $\Pi^{<,>}$ create the collision integral in the statistical equation, giving rise to {\em collisional} and {\em coherence damping}, the decay rate $\Gamma$ gives a finite width to pole functions, representing {\em Landau damping}. Due to their physically different roles one can consistently employ different approximations for these functions. In particular, one can adopt the spectral limit $\Gamma\to 0$ in pole equations while retaining $\Pi^{<,>}$ in the collision integral. This is of course exactly what one does to derive the standard Boltzmann limit.

The Hermitian self-energy function $\Pi_{\rm H}$ gives rise to modified dispersion relations due to background corrections, and keeping $\Pi_{\rm H}$ while neglecting $\Gamma$ corresponds to the usual quasiparticle approximation. This is consistent with the coupling constant expansion in our setup, where $\Pi_{\rm H}$ appears at the one-loop level and $\Gamma$ arises first at two loops. It is similarly consistent to use one-loop result for $\Pi_{\rm H}$ while computing $\Pi^{<,>}$ at two loops. This is a great simplification, because with the momentum independent one-loop self-energy the 2PI-renormalization procedure leads to a simple gap-equation for the effective mass~\cite{Kainulainen:2021eki,Kainulainen_2023}, but this would generalize to an integral equation for a momentum dependent $\Pi_{\rm H}$ beyond one loop. The small increase in accuracy this in most cases would bring, would come with a significant increase in computational load.

The decoupling process also removes the $\Pi^< \otimes \Delta_{\rm H}$-term from the statistical equation~\cref{eq:kadanoff_baym}. To see this, divide the full statistical function into a background and perturbation: $\Delta^< = \Delta^<_{\rm bg} + \delta\Delta^<$. Since the background solution by definition must make collision integral vanish, it must satisfy the equation
\begin{equation}
(\Delta^{-1}_0 - \Pi_{\rm H}) \otimes \Delta^<_{\rm bg} = \Pi^< \otimes \Delta_{\rm H}.
\label{eq:bg-equation}
\end{equation}
The definition of the background $\Delta^<_{\rm bg}$ thus entirely absorbs the $\Pi^< \otimes \Delta_{\rm H}$-term. Only background is thus directly coupled to the pole equations and it should be solved simultaneously with them. Indeed, from the pole equations we can derive an equation for the spectral function
\beq
(\Delta^{-1}_0 - \Pi_{\rm H}) \otimes {\cal A}  = \Gamma \otimes \Delta_{\rm H}.
\label{eq:spec-equation}
\eeq
Comparing~\cref{eq:bg-equation} and~\cref{eq:spec-equation}, we see that $\Pi^<$ has the same role in background equation that $\Gamma$ has in pole equations and it should be put to zero in the spectral limit $\Gamma \to 0$. The final observation we make is, that in the spectral limit~\cite{Herranen:2008di,Herranen:2010mh,Fidler:2011yq} the background and perturbation solutions can be combined, and the statistical equation for the full correlation function reduces to
\begin{equation}
  (\Delta^{-1}_0 - \Pi_{\rm H}) \otimes \Delta^< = \langle C \rangle.
\label{eq:kadanoff_baym_2}
\end{equation}
This equation, together with~\cref{eq:general-self-energy} and~\cref{eq:collision_integral}, is the starting point of our analysis.

%%%%%%%%%%%%%%%%%%%%%%%%%%%%%%%%%%%%%%%%%%%%%%%%%%%%%%%%%%%%%%%%%%%%%%%%%%%%%%%%%%%%%%%%%%%%%%%%%%%
\subsection{One-loop renormalization}
%%%%%%%%%%%%%%%%%%%%%%%%%%%%%%%%%%%%%%%%%%%%%%%%%%%%%%%%%%%%%%%%%%%%%%%%%%%%%%%%%%%%%%%%%%%%%%%%%%%
%
The 2PI-renormalization procedure for our setup was explained in detail in~\cite{Kainulainen_2023} in the Hartree approximation. These results are adequate here, because we are computing the Hermitian self-energy function $\Pi_{\rm H}$ to one-loop order. Moreover the 2-loop self-energy functions $\Pi^{<,>}$ entering collision integral are non-divergent, and since the collision integral vanishes in the background assumed in sensible renormalization conditions, the collision integral does not affect renormalization at all.  We can thus take the renormalized equations from~\cite{Kainulainen_2023} and simply add the finite collision integral on renormalized equations by hand\footnote{Because we are not performing a full self-consistent 2-loop renormalization in this paper, the coupling constant in the collision integral is not precisely defined. We identify it with $\lambda$ in equation~\cref{eq:physical_coiupling}, since this is the coupling appearing also in our one-loop self-energy term. Numerically, the difference between the different renormalized 2PI-couplings is small in the perturbative regime.}. 

We have not been explicit in denoting our variables and parameters being the bare ones so far. The 2PI-renormalization procedure is somewhat complicated, but the final equations can be written in an astonishingly simple form in terms of renormalized variables and couplings. To keep notation simple, we denote the renormalized quantities with the same names that we used above for their bare counterparts. The coupled equations for the renormalized background field $\bar\sigma$ and the renormalized 2-point function then become
\begin{align}
    (Z_2\square_{x} + M_\text{eff}^2) \bar\sigma &= \frac{1}{3}\lambda \bar\sigma^3,
    \label{eq:evo_equation_3}
\\
    (\square_{x} + M_\text{eff}^2) \Delta^<   &= -\langle C \rangle.
    \label{eq:hartree_equation_3}
\end{align}
The effective mass in these equations is solved from the gap equation
\begin{equation}
    M_\text{eff}^2 = m_{\rm eff}^2(t) + \frac{1}{2}\lambda\bar\sigma^2 
    + \frac{1}{2} \lambda(\Delta_{{\rm F}0}(M_\text{eff}^2) + \delta\Delta_{\rm F}),
\label{eq:gappi-1}
\end{equation}
where the finite background part of the local two-point function $\Delta_{\rm F0}$ depends on $M_\text{eff}^2$: 
\begin{align}
\Delta_{\rm F0}(M_\text{eff}^2) = \frac{\lambda}{32\pi}\left[ M^2_{\text{eff}}\ln\left( \frac{M^2_{\text{eff}}}{a^2 m^2_\chi}\right) - M^2_{\text{eff}} + a^2 m^2_\chi \right].
\end{align}
The finite non-equilibrium part $\delta\Delta_{\rm F}$ is a dynamical quantity, evaluated during the solution of the coupled equations. 

We have included a finite wave-function renormalization factor in the field equation~\cref{eq:evo_equation_3}, which was missed in~\cite{Kainulainen_2023}: 
\beq
Z_2 = 1 + \frac{\lambda}{64\pi^2}.
\eeq
The derivation of this result can be found in~\cite{Banik_Kainulainen_2026}. Let us finally note that the coupling $\lambda$ is related to the four-point function of the theory at zero external momenta $\lambda_{\rm R} \equiv \Gamma^{(4)}(0)$ by
\begin{equation}
\lambda = \frac{32\pi^2}{9}\left[ \Big(1+ \frac{27\lambda_{\rm R}}{32\pi^2} \Big)^{1/3} -1 \right]
\label{eq:physical_coiupling}
\end{equation}
This is just the 2PI-Hartree equivalent of the rule that relates couplings in the usual Coleman-Weinberg effective potential approach: $\lambda_{\rm R} = \lambda_{\rm CW} + 9\lambda^2_{\rm CW}/32\pi^2$, where extra $\lambda^2_{\rm CW}$-factor arise from the finite triangle and box-contributions to $\Gamma^{(4)}(0)$. Inverting this relation gives a formula similar to~\cref{eq:physical_coiupling}: $\lambda_{\rm CW} = (16\pi^2/9)[(1+9\lambda_{\rm R}/(8\pi^2))^{1/2}-1]$.

%%%%%%%%%%%%%%%%%%%%%%%%%%%%%%%%%%%%%%%%%%%%%%%%%%%%%%%%%%%%%%%%%%%%%%%%%%%%%%%%%%%%%%%%%%%%%%%%%%%
\subsection{Moment equations}
%%%%%%%%%%%%%%%%%%%%%%%%%%%%%%%%%%%%%%%%%%%%%%%%%%%%%%%%%%%%%%%%%%%%%%%%%%%%%%%%%%%%%%%%%%%%%%%%%%%
%
The next step in the derivation is moving to the Wigner space, assuming spatial homogeneity, and taking the two lowest frequency moments of Wigner-transformed equation~\cref{eq:hartree_equation_3}. This procedure was also described in detail in~\cite{Kainulainen_2023} (following~\cite{Herranen:2010mh}) and we simply quote the final result\footnote{We correct a typo in~\cite{Herranen:2010mh}, where the sign of the $\text{Im} \langle C_{0\bm k} \rangle$ term was incorrect.}: 
\begin{align}
    \frac14 \partial_t^2 \rho_{0\bm k} + \omega_{\bm k}^2 \rho_{0\bm k} - \rho_{2\bm k} 
    &= \text{Im} \langle C_{0\bm k} \rangle 
    \label{eq:rho0_eom}  \\
    \partial_t \rho_{1\bm k} &= \text{Re}\langle C_{0\bm k} \rangle 
    \label{eq:rho1_eom}, \\
    \partial_t \rho_{2\bm k} - \frac12 \partial_t \left[ M_\text{eff}^2 \right] \rho_{0\bm k} 
    &= \text{Re}\langle C_{1\bm k} \rangle,
    \label{eq:rho2_eom}
\end{align}
where the moment functions are defined as
\begin{equation}
    \rho_{\alpha\bm k}(t) \equiv \int \frac{dk_0}{2\pi} k_0^\alpha i\Delta^<_{\bm k}(k_0,t)
    \label{eq:moment_def_stable}
\end{equation}
and renormalized effective mass $M_\text{eff}^2$ is explicitly\footnote{Note the difference of $\lambda/4 \to \lambda/4!$ in the normalization of the coupling in comparison with~\cite{Kainulainen_2023}.}
\begin{equation}
    M^2_{\text{eff}} = m^2_{\rm eff}(t) + \frac12 \lambda \bar\sigma^2 
    + \frac12 \lambda\Delta_{\rm F0}
    + \frac12 \lambda \int_{\bm k}\left( \rho_{0\bm k} - \frac{\theta(\omega^2_{\bm k})}{2\omega_{\bm k}} \right). 
\end{equation} 
The energy in~\cref{eq:rho2_eom} is defined as $\omega_{\bm k}^2 = {\bm k}^2 + M_\text{eff}^2$. These equations, coupled with the equation to the 1-point function, constitute an initial value problem, and the initial conditions suitable for our example setup have been discussed in~\cite{Kainulainen_2023}.
Finally, the moments of the collision integral are denoted as $\langle C_{\alpha {\bm k}}(t)\rangle$, and they can be computed as
\begin{equation}
\begin{split}
    \langle C_{\alpha \bm k} (t) \rangle  
    & = \lim\limits_{r_0 \to 0}
    \int \frac{{\rm d}k_0}{2\pi} k_0^\alpha C_{\bm k}(k_0,t)e^{-ik_0r_0}
\\
    & = \lim\limits_{r_0 \to 0}\frac12\int {\rm d}w^0 (i\partial_{r^0})^\alpha  \left[\Pi_{{\bm k}}^> \left(t + \frac{r^0}{2},w^0 \right) \Delta^<_{{\bm k}} \left( w^0, t - \frac{r^0}{2} \right) \right] - (> \leftrightarrow<),
    \label{eq:CI_moments_partially_integrated}
 \end{split}
\end{equation}
where the second line assumes vanishing of the boundary terms. Note that the collision integral is dependent on the full two-point function in the two-time representation, both directly and through the self-energy. In contrast, the moments $\rho_\alpha(t)$ are functions of the equal-time correlation function $\Delta^<(t,t)$ and its few lowest relative-time direction time derivatives. These clearly do not contain sufficient information to reconstruct the full non-local correlation function, and further approximations are needed to evaluate the collision integrals.

%%%%%%%%%%%%%%%%%%%%%%%%%%%%%%%%%%%%%%%%%%%%%%%%%%%%%%%%%%%%%%%%%%%%%%%%%%%%%%%%%%%%%%%%%%%%%%%%%%%%
%%%%%%%%%%%%%%%%%%%%%%%%%%%%%%%%%%%%%%%%%%%%%%%%%%%%%%%%%%%%%%%%%%%%%%%%%%%%%%%%%%%%%%%%%%%%%%%%%%%%%
%
\section{The phase-space of the Wightman functions, stable branch}
\label{sec:ASA_approximation}
%
%%%%%%%%%%%%%%%%%%%%%%%%%%%%%%%%%%%%%%%%%%%%%%%%%%%%%%%%%%%%%%%%%%%%%%%%%%%%%%%%%%%%%%%%%%%%%%%%%%%%%
%%%%%%%%%%%%%%%%%%%%%%%%%%%%%%%%%%%%%%%%%%%%%%%%%%%%%%%%%%%%%%%%%%%%%%%%%%%%%%%%%%%%%%%%%%%%%%%%%%%%%

In order to find a tractable approximation for the collision integral, we study the phase-space structure of the two-point function first in the collisionless limit. We keep the Hermitian self-energy correction, however, which accounts for the background corrections on dispersion relations. This problem was studied in~\cite{Herranen:2008di,Herranen:2010mh,Fidler:2011yq,Herranen:2011zg}, where it was shown to give rise to spectral solutions including zero-frequency shells that carry non-local coherence information. In these papers, the authors worked in the frequency space, but we can obtain the same results also working in the direct space. However, since we are interested in homogeneous problems, we start from the collisionless~\cref{eq:hartree_equation_3} Fourier transformed with respect to spatial coordinates:
\begin{equation}
    (\partial^2_{u^0} + \omega_{\bm k}^2) i\Delta_{\bm k}^<(u^0,v^0) = 0,
    \label{eq:hartree_equation}
\end{equation}
where $\omega^2_{\bm k} = {\bm k}^2 + M_\text{eff}^2$. The same equation is obeyed by $i\Delta_{\bm k}^>$ as well. While the time-evolution of the effective mass is very non-trivial, equation~\cref{eq:hartree_equation} is formally just the familiar Klein-Gordon equation. In this chapter we will assume that ${\bm k}^2 > M_\text{eff}^2$. This includes both normal excitations with $M_\text{eff}^2>0$ and stable tachyonic solutions with $M_\text{eff}^2<0$ but ${\bm k}^2 + M_\text{eff}^2>0$. The latter still behave as propagating waves, only with a modified dispersion relation. The unstable tachyonic branch with ${\bm k}^2 + M_\text{eff}^2<0$ requires special care and will be treated separately.

If $M_{\rm eff}^2$ remains constant over the timescale under consideration, the solution to~\cref{eq:hartree_equation} can be written as $i\Delta^s_{\bm k} = \sum_a e^{-ia\omega_{\bm k}(u_0-t)}K_{\bm k}^{s,a}(t,v_0)$, representing free evolution from some initial time $t$ with some arbitrary coefficient functions $K_{\bm k}^{s,a}(t,v_0)$. However, the Hermiticity property $i\Delta^s(u_0,v_0) = (i\Delta^s(v_0,u_0))^*$ further constrains this solution to the form
\begin{align}
\nonumber \\[-15pt]
i \Delta^{s}_{\bm k}(u^0,v^0;t) &= \sum_{a,b} e^{-ia\omega_{\bm k}(u^0-t)} i\Delta^{s,ab}_{\bm k}(t,t) e^{-ib\omega_{\bm k}(t - v^0)}, 
\label{eq:hartree_solution}
\end{align}
\vskip-4pt
\noindent
where $i\Delta^{s,ab}_{\bm k}(t,t)$ are some yet unknown components of the local correlation function at time $t$, satisfying $i\Delta^{s,ab}_{\bm k} =(i\Delta^{s,ab}_{\bm k})^*$. Taking the Wigner transform of~\cref{eq:hartree_solution} immediately gives the spectral coherent quasiparticle (cQPA) solution derived in~\cite{Herranen:2010mh}: 
\begin{align}
\nonumber \\[-13pt]
i\Delta^s_{\bm k}(k_0,x_0) = 2\pi \sum_{ab} i\Delta^{s,ab}_{\bm k}(t,t) 
e^{-2i\Delta\omega^{ab}_{\bm k}(t-x_0)} \delta(k_0 - \bar\omega^{ab}_{\bm k}).
\label{eq:on_shell_wigner}
\end{align}
\vskip-4pt
\noindent
Here $x_0 \equiv (u_0+v_0)/2$, $\Delta\omega^{ab}_{\bm k} \equiv (a-b)\omega_{\bm k}/2$ and $\bar \omega^{ab}_{\bm k} \equiv (a+b)\omega_{\bm k}/2$. We will also frequently use the notation $\omega^a_{\bm k} \equiv a\omega_{\bm k}$. Our approach is to use solutions~\cref{eq:hartree_solution,eq:on_shell_wigner} as an Ansatz for the full two-point function \emph{only inside the collision integral}. The error made in neglecting the effect of collisions on this approximation are higher order in the collision term for the full evolution. The reference time variable $t$ should be chosen such that the coefficient functions correspond to known values of the two-point function. A natural choice in our application is the external time of equations~\cref{eq:rho0_eom,eq:rho2_eom}. The simple form of~\cref{eq:hartree_solution} provides a natural interpretation for this approximation: particles are (or more precisely the local density operator is) assumed evolve freely between the point $(t,t)$ and the (decohering) collision point at $(u_0,v_0)$.

\begin{figure}[t]
   \centering
   \includegraphics[width=0.35\linewidth]{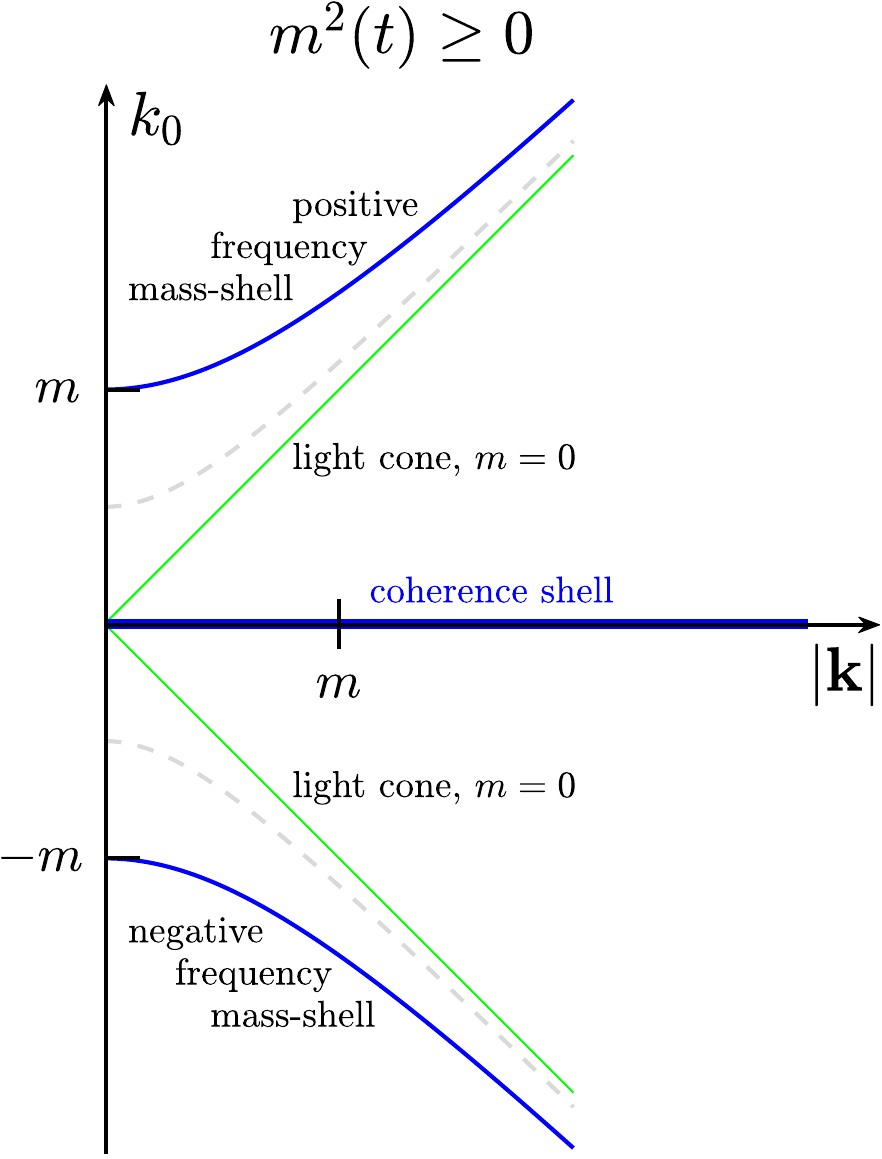} 
   \;\;\;\;\;
   \includegraphics[width=0.36\linewidth]{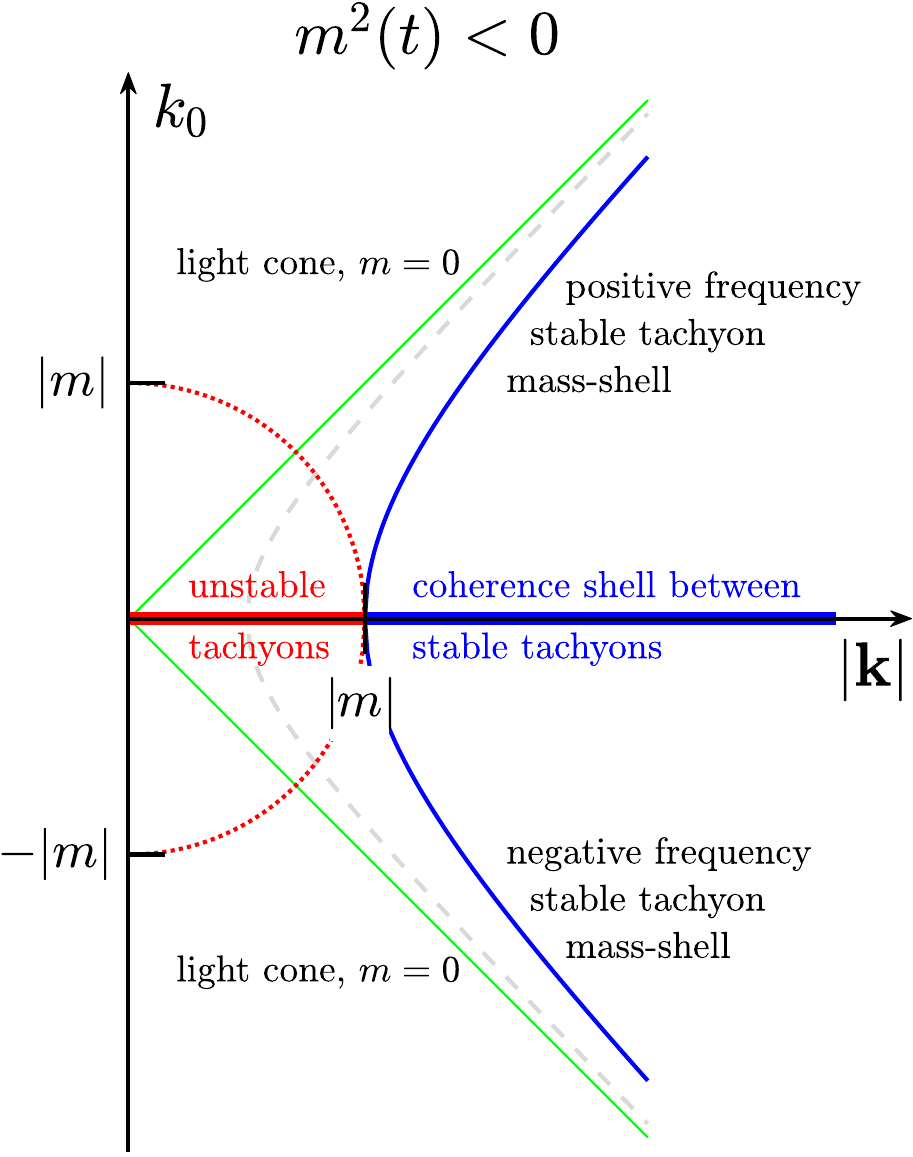} 
\vskip0.2truecm
\caption{Shown are the cQPA-dispersion relations in the normal branch with $m^2(t) \ge 0$ (left panel) and in the tachyonic branch $m^2(t) < 0$ (right panel). On the left panel the line at $k_0=0$ hosts the coherence shells. On the right panel the entire phase space has fallen to spacelike region and the coherence shell at $k_0=0$ divides into stable and unstable branches. The red dotted circle show the positions of the poles in complex frequency ($-i(k_0)_{\rm pole}$) for the convergent unstable tachyons (see section~\cref{sec:ASA_approximation_unstable}).}
\label{fig:dispersion_relations}
\end{figure}
\vskip0.4truecm

For $M_{\rm eff}^2>0$, the spectral solution~\cref{eq:on_shell_wigner} splits into three different shells: two \emph{mass-shells} at $k_0 = \pm \omega_{\bm k}$ familiar from thermal field theory and a \emph{coherence shell} at $k_0 = 0$, which contains information on the correlation between positive and negative frequency excitations. This structure is depicted on the left panel of figure~\cref{fig:dispersion_relations}. The mass-shell solutions approach the light cone as $M_{\rm eff}^2$ tends to zero, and pass through it as $M_{\rm eff}^2$ turns negative. The tachyonic branch with $M_{\rm eff}^2<0$, shown in the right panel of figure~\cref{fig:dispersion_relations}, further divides into a stable branch with ${\bm k}^2 + M_{\rm eff}^2>0$ and an unstable one with ${\bm k}^2 + M_{\rm eff}^2<0$. The former shows a similar structure to the non-tachyonic branch, except that the tachyonic mass- and coherence shells converge to $k_0=0$ when ${\bm k}^2 + M_{\rm eff}^2 \to 0$. This branch can nevertheless be described by the arguments given in this section. The unstable branch requires more attention and will be discussed later. 

%%%%%%%%%%%%%%%%%%%%%%%%%%%%%%%%%%%%%%%%%%%%%%%%%%%%%%%%%%%%%%%%%%%%%%%%%%%%%%%%%%%%%%%%%%%%%%%%%%%%%%%%
\subsection{Dimensionless shell-functions}
%%%%%%%%%%%%%%%%%%%%%%%%%%%%%%%%%%%%%%%%%%%%%%%%%%%%%%%%%%%%%%%%%%%%%%%%%%%%%%%%%%%%%%%%%%%%%%%%%%%%%%%%

From now on, we will refer to the coefficients $\Delta^{s,ab}_{\bm k}$ as shell functions. To make a connection with the notation of~\cite{Herranen:2008di,Herranen:2010mh,Fidler:2011yq} and to obtain dimensionless functions, we scale them with energy, defining:
\begin{equation}
i\Delta^{s,ab}_{\bm k} \equiv (a\delta_{a,b} + \delta_{a,-b})\frac{f^{s,ab}_{\bm k}}{2\omega_{\bm k}}      \equiv \frac{\hat f^{s,ab}_{\bm k}}{2\omega_{\bm k}}.
\label{eq:fhat-functions}
\end{equation}
The signs for the hat-less scaled mass-shell functions $f^{s,\pm\pm}_{\bm k}$ are now such that they coincide with the standard momentum distribution functions from kinetic theory in the Boltzmann limit. The shell functions are easily related to the moments at the time $x^0 = t$:
\begin{align}
    \rho_{0\bm k} &= \frac{1}{2\omega_{\bm k}}
     \left(f^{<++}_{\bm k} - f^{<--}_{\bm k} + f^{<+-}_{\bm k} + f^{<-+}_{\bm k} \right), 
    \label{eq:rho0_cqpa} \\
    \rho_{1\bm k} &= \;\frac{1}{2} \left( f^{<++}_{\bm k} + f^{<--}_{\bm k} \right), 
    \label{eq:rho1_cqpa} \\
    \rho_{2\bm k} &= \frac{\omega_{\bm k}}{2} \left(f^{<++}_{\bm k} - f^{<--}_{\bm k} \right), 
    \label{eq:rho2_cqpa} \\
    \partial_t \rho_{0\bm k} &\approx -\frac{i}{2} \left(f^{<+-}_{\bm k} - f^{<-+}_{\bm k} \right). \label{eq:rho0p_cqpa}
\end{align}
To get the last equation, we have used the lowest order (in gradients and couplings) equation of motion for the shell-functions, 
\begin{equation}
\partial_t f_{\bm k}^{s,ab} \approx -2i\Delta\omega_{\bm k}^{ab} f_{\bm k}^{s,ab},
\label{eq:eom_for_fs}
\end{equation}
which is easily derived from moment equations~\cref{eq:rho0_eom,eq:rho1_eom,eq:rho2_eom} in the collisionless limit~\cite{Herranen:2010mh}. 

In practice, the approximation~\cref{eq:hartree_solution} allows us to write the collision integrals in terms of the shell functions $f^{s,ab}_{\bm k}$, and for that we will need the inverses of the relations~\cref{eq:rho0_cqpa,eq:rho1_cqpa,eq:rho2_cqpa,eq:rho0p_cqpa}. Explicitly, these relations read~\cite{Herranen:2010mh}:
\begin{align}
    f^{<\pm\pm}_{\bm k} &= \rho_{1\bm k} \pm \frac{1}{\omega_{\bm k}}\rho_{2\bm k}
    \label{eq:fpmpm}\\
    f^{<\pm\mp}_{\bm k} &= 
    \omega_{\bm k}\rho_{0\bm k} - \frac{1}{\omega_{\bm k}}\rho_{2\bm k}
    \pm \frac{i}{2}\partial_t\rho_{0\bm k}.
    \label{eq:fpmmp}
\end{align}
An analogous reduction can introduced for the function $i\Delta^>_{\bm k}$. Moreover, from the condition $i\Delta^>_{\bm k}-i\Delta^<_{\bm k} = 2{\cal A}_{\bm k}$, one finds that the shell functions $f^{>ab}_{\bm k}$ are related to $f^{<ab}_{\bm k}$ by 
\begin{equation}
f^{>ab}_{\bm k} = \delta_{ab} + f^{<ab}_{\bm k}.
\label{eq:useful_property_0}
\end{equation}
There are still a few more useful relations between the shell functions. First, from relation $i\Delta^>(u,v) = i\Delta^<(v,u) \Leftrightarrow i\Delta^>(k,x) = i\Delta^<(-k,x)$, we get
\begin{equation}
{\hat f}^{>ab}_{{\bm k}} = {\hat f}^{<-b-a}_{-{\bm k}},
\label{eq:useful_property_1}
\end{equation}
which implies that $f_{-\bm k}^{>\mp\mp}=-f_{\bm k}^{<\pm\pm}$ and $f_{-\bm k}^{>\pm\mp}=f_{\bm k}^{<\pm\mp}$. This condition also shows, that real scalar particles are their own antiparticles, $\bar f^{<++}_{\bm k} = -f^{>--}_{-{\bm k}} = f^{<++}_{{\bm k}}$, where the first equality is the Feynman-Stueckelberg relation and the last followed from~\cref{eq:useful_property_1}. Second, from Hermiticity $(i\Delta^s(u,v))^\dagger = i\Delta^s(v,u) \Leftrightarrow (i\Delta^s(k,x))^\dagger = i\Delta^s(k,x)$, we find
\begin{equation}
f^{<\pm\pm}_{{\bm k}} \in {\mathrm R} \quad {\rm and} \quad 
(f^{s+-}_{{\bm k}})^* = f^{s-+}_{{\bm k}}.
\label{eq:useful_property_2}
\end{equation}
This shows that functions $f^{<ab}_{\bm k}$ contain exactly four independent degrees of freedom. Third, in what follows we can often freely sum over one of the indices, and so it is useful to define the summed quantities: 
\begin{equation}
f^{s\pm}_{{\bm k}} \equiv \sum_a \hat f^{sa\pm}_{\bm k} 
= \pm f^{s\pm\pm}_{\bm k} + f^{s\mp\pm}_{\bm k}.
\label{eq:summed_fs}
\end{equation}
\vskip-0.2cm\noindent
Using~\cref{eq:useful_property_1,eq:useful_property_2}, one can show that $f^{<-}_{{\bm k}} = (f^{>+}_{-{\bm k}})^*$. When summing over the second index, we get $\sum_a \hat f^{s\pm a}_{\bm k} = (f^{s\pm}_{\bm k})^*$ instead. 

The dimensionless shell-functions are very useful for the discussion of the stable branch modes and as stated above, they are normalized such that they reduce to the standard Bose-Einstein distribution in thermal limit: $f^{<++}_{\bm k}\to f_{{\rm BE},{\bm k}}(T)$. Shell functions can be defined also for the unstable tachyonic modes such that relations~\cref{eq:rho0_cqpa,eq:rho1_cqpa,eq:rho2_cqpa,eq:rho0p_cqpa,eq:fpmpm,eq:fpmmp} and the properties derived above hold with the analytic continuation $\omega_{\bm k}\to i|\omega_{\bm k}|$, apart from the complex conjugation rule~\cref{eq:useful_property_2}, which will be discussed in section~\cref{subsec:unstable_tarchyon_collisions}.

%%%%%%%%%%%%%%%%%%%%%%%%%%%%%%%%%%%%%%%%%%%%%%%%%%%%%%%%%%%%%%%%%%%%%%%%%%%%%%%%%%%%%%%%%%%%%%%%%%%%%%%
%%%%%%%%%%%%%%%%%%%%%%%%%%%%%%%%%%%%%%%%%%%%%%%%%%%%%%%%%%%%%%%%%%%%%%%%%%%%%%%%%%%%%%%%%%%%%%%%%%%%%%%
%
\section{The collision integral, stable branch}
\label{sec:collision_integral}
%
%%%%%%%%%%%%%%%%%%%%%%%%%%%%%%%%%%%%%%%%%%%%%%%%%%%%%%%%%%%%%%%%%%%%%%%%%%%%%%%%%%%%%%%%%%%%%%%%%%%%%%%
%%%%%%%%%%%%%%%%%%%%%%%%%%%%%%%%%%%%%%%%%%%%%%%%%%%%%%%%%%%%%%%%%%%%%%%%%%%%%%%%%%%%%%%%%%%%%%%%%%%%%%%

For a concrete example, we will next use the approximation scheme of the previous section to compute the lowest order contribution to the collision integral in our setup. Before considering explicit self-energies however, we note that inserting the Ansatz~\cref{eq:hartree_solution} for $\Delta^{<,>}_{\bm k}$ along with the definition~\cref{eq:fhat-functions}, one can write the collision integral moments~\cref{eq:CI_moments_partially_integrated} as
\begin{equation}
    \langle C_{\alpha \bm k} (t) \rangle  
    = - \frac{1}{4\omega_{\bm k}}\sum_{ab} 
        \left[ \left(\frac{i}{2}\partial_t + \delta_{ab}\omega^a_{\bm k} \right)^\alpha 
               i\Pi^s_{{\bm k},\rm out}(\omega^a_{\bm k}) \right] \hat f^{<ab}_{\bm k}  
      - (> \leftrightarrow <),
    \label{eq:CI_moments_partially_integrated_2}
\end{equation}
where
\begin{equation}
  i\Pi^s_{{\bm k},\rm out}(\omega^a_{\bm k}) 
          \equiv \int {\rm d}w_0 i\Pi_{\bm k}^s(t,w_0) e^{-i\omega^a_{\bm k}(t-w_0)}
\label{eq:Pi-in}
\end{equation}
are generic functions that have a central role in what follows. From the general conditions $(i\Pi^<(k,x))^\dagger = i\Pi^<(k,x) = i\Pi^>(-k,x)$, it follows that
\begin{equation}
  i\Pi^<_{{\bm k},\rm out}(-\omega^a_{\bm k}) = (i\Pi^>_{-{\bm k},\rm out}(\omega^a_{\bm k}))^*.
 \label{eq:hermit-cond-pi-out} 
\end{equation}

Only the the two lowest moments are needed in our evolution equations~\cref{eq:rho0_eom,eq:rho1_eom,eq:rho2_eom}. The zeroth moment can be written as
\begin{align}
    \langle C_{0{\bm k}}(t)\rangle  
     &= -\frac{1}{4\omega_{\bm k}} \sum_{ab} \left( i\Pi^>_{{\bm k},\rm out}(\omega^a_{\bm k})
                  \hat f^{<ab}_{\bm k} - (> \leftrightarrow <) \right)
    \nonumber\\
     &= -\frac{i}{2\omega_{\bm k}} {\rm Im} \left( i\Pi^>_{{\bm k},\rm out}(\omega_{\bm k})
                  (f^{<+}_{\bm k})^* - (> \leftrightarrow <) \right).
\label{eq:zeroth_moment}
\end{align}
In the second line we summed over frequencies, making use of~\cref{eq:hermit-cond-pi-out} and the properties of shell-functions~\cref{eq:useful_property_1,eq:useful_property_2,eq:summed_fs}. The result is purely imaginary as expected. For a real scalar field the first moment is directly related to the canonical commutation relation giving $\rho_{1{\bm k}}=-1/2$, and the condition Re$\langle C_{0\bm k}\rangle=0$ reduces~\cref{eq:rho1_eom} to $\partial_t \rho_{1{\bm k}} = 0$, which ensures the conservation of $\rho_{1{\bm k}}$. The first moment requires only a little more work. We can write it as
\begin{align}
   \langle C_{1{\bm k}}(t)\rangle  
       &= -\frac{1}{4\omega_{\bm k}} \sum_{a} 
        \Big( \omega_{\bm k}i\Pi^>_{{\bm k},\rm out}(\omega^a_{\bm k}) f^{<aa}_{\bm k} 
      + \frac{i}{2}\big(\partial_t i\Pi^>_{{\bm k},\rm out}(\omega^a_{\bm k})\big) (f^{<a}_{\bm k})^* - (> \leftrightarrow <) \Big).
      \nonumber \\
      &= -\frac{1}{2\omega_{\bm k}} {\mathrm Re}
        \Big( i\Pi^>_{{\bm k},\rm out}(\omega_{\bm k}) \omega_{\bm k}f^{<++}_{\bm k} 
      + \frac{i}{2}\big(\partial_t i\Pi^>_{{\bm k},\rm out}(\omega_{\bm k})\big) (f^{<+}_{\bm k})^* - (> \leftrightarrow <) \Big),
\label{eq:first_moment}
\end{align}
where we again used~\cref{eq:hermit-cond-pi-out} to get rid of the frequency sums in the first line. This time the result turns out to be purely real. Until now, all formulas are valid to any order in the perturbative expansion. Next, we introduce our two-loop approximation.

%===================================================================================================
%===================================================================================================
%
% Note: feynmf-version of the diagrams commented out.
\begin{figure}[t]
   \centering
   \includegraphics[width=0.6\linewidth]{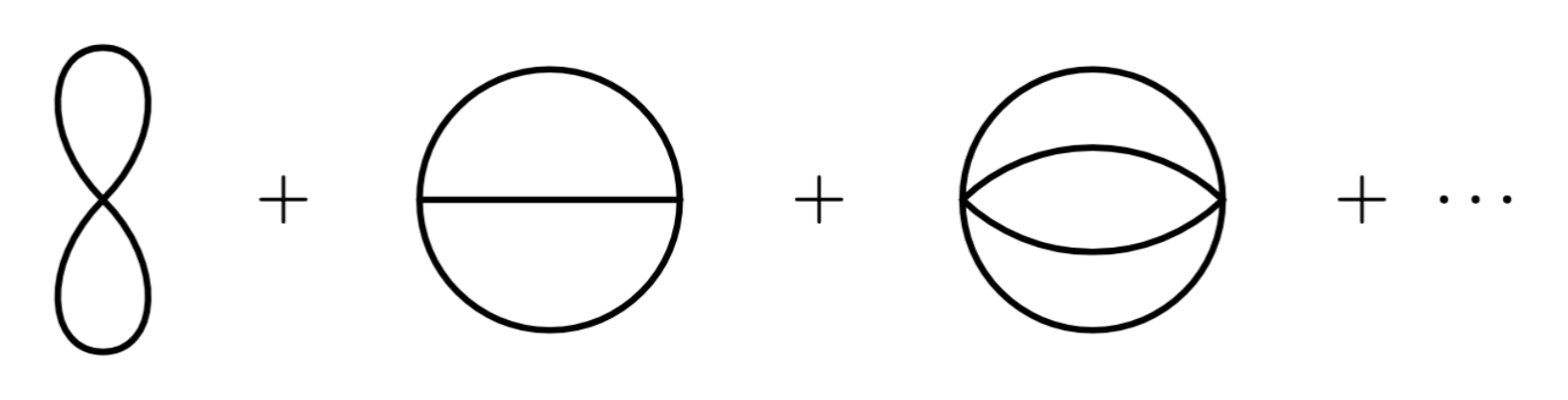} 
%\begin{fmffile}{diagram}
%\begin{equation*}
%\hspace{-1em}
%\parbox{17mm}
%{
%\begin{fmfgraph*}(70,40)
%\fmfleft{i}
%\fmfright{o}
%\fmf{phantom}{i,v,v,o}
%\fmf{plain}{v,v}
%\fmf{plain,left=90}{v,v}
%\end{fmfgraph*}
%}
%\hspace{0.7em}
%+
%\hspace{1.3em}
%\parbox{20mm}
%{
%\begin{fmfgraph*}(40,40)
%\fmfleft{v1}
%\fmf{plain,left,tension=.3}{v1,v2,v1}
%\fmf{plain}{v1,v2}
%\fmfright{v2}
%\fmf{phantom}{v1,v3,v2}
%\end{fmfgraph*}
%}
%\hspace{-0.2em}
%+
%\hspace{1.4em}
%\parbox{20mm}
%{
%\begin{fmfgraph*}(40,40)
%\fmfleft{v1}
%\fmf{plain,left,tension=.3}{v1,v2,v1}
%\fmf{plain,left=0.4}{v1,v2,v1}
%\fmfright{v2}
%\fmf{phantom}{v1,v3,v2}
%\end{fmfgraph*}
%}
%\hspace{-0.2em} +\hspace{0.4em} \cdots
%\end{equation*}
%\end{fmffile}
%\vspace{0em}
\caption{The 2-particle irreducible diagrams contributing to $\Gamma_2$ up to order $\lambda^2$.}
\label{fig:gamma2_expansion}
\end{figure}
%===================================================================================================
%===================================================================================================

%%%%%%%%%%%%%%%%%%%%%%%%%%%%%%%%%%%%%%%%%%%%%%%%%%%%%%%%%%%%%%%%%%%%%%%%%%%
\paragraph{Two-loop expansion for self energies.}
%%%%%%%%%%%%%%%%%%%%%%%%%%%%%%%%%%%%%%%%%%%%%%%%%%%%%%%%%%%%%%%%%%%%%%%%%%%

The three lowest order vacuum bubble diagrams contributing to $\Gamma_2$ are shown in figure~\cref{fig:gamma2_expansion}. The first diagram is the Hartree term, which defines the Hermitian self-energy function in our approximation, but does not contribute to the collision integral. The lowest order contribution to the latter come from the last two $\lambda^2$-diagrams shown. The self-energy arising from the middle diagram reads
\begin{equation}
    i\Pi^{ab}_{B}(u,v) = \frac{\lambda^2}{2} \sigma(u)\sigma(v) (i\Delta^{ab}(u,v))^2.
\label{eq:self_energy_B}
\end{equation}
This diagram invokes energy exchange from the dynamical one-point function in three-body scattering processes, and its contribution can be significant in a general setup. However, the one-point function in the setup of~\cite{Fairbairn:2018bsw}  stays small both in the Hartree calculation of~\cite{Kainulainen_2023} and in the lattice implementation of~\cite{Kainulainen:2024etd}, which will suppress this contribution by several orders of magnitude compared to the other diagrams. As such, we will neglect~\cref{eq:self_energy_B} from now on.  

The dominant contribution to the collision integral arises from the last diagram in figure~\cref{fig:gamma2_expansion}. Assuming homogeneity and moving to mixed space with respect to spatial coordinates, this diagram corresponds to
\begin{equation}
    i\Pi^>_{C{\bm k}}(t,w_0) = \frac{\lambda^2}{6} \int_{{\bm k}_1{\bm k}_2{\bm k}_3} 
    i\Delta^>_{{\bm k}_1}(t,w_0) i\Delta^>_{{\bm k}_2}(t,w_0) i\Delta^>_{{\bm k}_3}(t,w_0) 
    (2\pi)^3 \delta^3({\bm k} - \Sigma_i {\bm k}_i),
\label{eq:Pi-greater}    
\end{equation}
where $\int_{\bm k} \equiv \int {\rm d}^3k/(2\pi)^3$. The self-energy $\Pi^<_{C{\bm k}}(t,w_0)$ can be obtained from~\cref{eq:Pi-greater} by exchanging $< \leftrightarrow >$. From now on we shall drop the diagram specification from the self-energy, with the understanding that functions $\Pi^{<,>}_{\bm k}$ refer to the diagram $C$. 

%%%%%%%%%%%%%%%%%%%%%%%%%%%%%%%%%%%%%%%%%%%%%%%%%%%%%%%%%%%%%%%%%%%%%%%%%%%%%%%%%%%%%%%%%%%%%%%%%%%%%%%
%\paragraph{Computing effective self-energy functions.}
%%%%%%%%%%%%%%%%%%%%%%%%%%%%%%%%%%%%%%%%%%%%%%%%%%%%%%%%%%%%%%%%%%%%%%%%%%%%%%%%%%%%%%%%%%%%%%%%%%%%%%%
%
Inserting the Ansatz~\cref{eq:hartree_solution} now also to propagators $i\Delta^>_{{\bm k}_i}(t,w_0)$ within $i\Pi_{\bm k}^>(t,w_0)$ given by~\cref{eq:Pi-greater}, we see that the $w_0$-integral in~\cref{eq:Pi-in} reduces to an integral over phases that can be immediately carried out:
\begin{equation}
\int {\rm d}w_0 e^{-i(\sum_i \omega^{b_i}_{{\bm k}_i}-\omega^a_{\bm k})(w_0-t)} = 
2\pi \delta({\textstyle\sum_i} \omega^{b_i}_{\bm k_i}-\omega^a_{\bm k}).
\label{eq:phase-integral-at-vertex}
\end{equation}
\vskip -3pt \noindent
This gives the nontrivial frequency signature dependent energy conservation at the internal vertex in the self-energy diagram. Note in particular how the explicit $t$-dependent phase vanishes in~\cref{eq:phase-integral-at-vertex}. Using this result, we can write~\cref{eq:Pi-in} as 
\begin{align}
  i\Pi^>_{{\bm k},\rm out}(\omega^a_{\bm k}) 
    &=  \frac{\lambda^2}{6} \sum_{a_i, b_i} \int {\rm dPS}^{\bm k}_{a,\{b_i\}} \; 
           {\hat f}^{>a_1 b_1}_{{\bm k}_1} 
           {\hat f}^{>a_2 b_2}_{{\bm k}_2} 
           {\hat f}^{>a_3 b_3}_{{\bm k}_3}
  \nonumber \\
    &=  \frac{\lambda^2}{6} \sum_{\{b_i\}} \int {\rm dPS}^{\bm k}_{a,\{b_i\}} \; 
           f^{>b_1}_{{\bm k}_1} 
           f^{>b_2}_{{\bm k}_2} 
           (f^{<b_3}_{{\bm k}_3})^*,
\label{eq:Pi-in_2}
\end{align}
\vskip-0.1truecm \noindent
where the phase space factor is defined as
\begin{equation}
\int {\rm dPS}^{{\bm k}}_{a,\{b_i\}} \equiv \int \prod_i \frac{{\rm d}^3 k_i}{(2\pi)^32\omega_{\bm k_i}}\, (2\pi)^4 \delta^{(3)} ({\bm k}_1 + {\bm k}_2 - {\bm k}_3-{\bm k}) \delta(\omega^{b_1}_{{\bm k}_1}
+\omega^{b_2}_{{\bm k}_2}-\omega^{b_3}_{{\bm k}_3} - \omega^a_{\bm k}).
\label{eq:phase_space}
\end{equation}
\vskip-0.1truecm \noindent
To get this symmetric form, we relabeled the dummy summation index $b_3\to -b_3$ and the integration variable ${\bm k}_3\to -{\bm k}_3$ and used the shell-function identities~\cref{eq:useful_property_1,eq:useful_property_2,eq:summed_fs} to derive the form in the second line of~\cref{eq:Pi-in_2}. The result for $i\Pi^>_{{\bm k},\rm out}(\omega^a_{\bm k})$ follows from~\cref{eq:Pi-in_2} by exchange $<\leftrightarrow >$. 

At this point, our two-loop results are still valid for both the normal branch excitations with $M_{\rm eff}^2>0$ and the stable tachyons. A difference between the two arises at the level of kinematical constraints, which allow a further simplification in the normal branch.

%%%%%%%%%%%%%%%%%%%%%%%%%%%%%%%%%%%%%%%%%%%%%%%%%%%%%%%%%%%%%%%%%%%%%%%%%%%%
\paragraph{Collision integrals for a positive mass squared function.}
%%%%%%%%%%%%%%%%%%%%%%%%%%%%%%%%%%%%%%%%%%%%%%%%%%%%%%%%%%%%%%%%%%%%%%%%%%%%

It is easy to see that for $M_{\rm eff}^2(t)>0$, kinematic constraints in the phase space integral~\cref{eq:phase_space} lead to the usual condition for frequency signs: $b_1+b_2 = b_3+a$. If we take $a=1$, we see there are exactly three possibilities $(b_1,b_2,b_3) = {(1,1,1),(1,-1,-1),(-1,1,-1)}$. A little algebra using the above relations between the various $f_{\bm k}$-functions shows that all triplets give the same result in~\cref{eq:Pi-in_2}, so that finally
\begin{equation}
  i\Pi^>_{{\bm k},\rm out}(\omega_{\bm k}) 
    =  \frac{\lambda^2}{2} \int {\rm dPS}_{\bm k} \;%_{1\{1\}} \; 
           f^{>+}_{{\bm k}_1} 
           f^{>+}_{{\bm k}_2} 
           (f^{<+}_{{\bm k}_3})^*,
\label{eq:Pi-in_4}
\end{equation}
where ${\rm dPS}_{\bm k} \equiv {\rm dPS}^{\bm k}_{1,\{1\}}$ is the standard phase space factor with all frequency variables set to +1 in~\cref{eq:phase_space}. The result for $\Pi^>_{{\bm k},\rm out}(\omega_{\bm k})$ can be treated similarly. Given~\cref{eq:Pi-in_4}, we can now compute the first two moments of the collision integral in the case of a positive mass squared. For the zeroth moment, we find: 
\begin{equation}
    \langle C_{0{\bm k}}(t)\rangle  
= -\frac{i\lambda^2}{4\omega_{\bm k}} \int {\rm dPS}_{\bm k}\;%_{1\{1\}} \; 
                 {\rm Im} \left(f^{>+}_{{\bm k}_1} 
                                f^{>+}_{{\bm k}_2} 
                               (f^{<+}_{{\bm k}_3})^*
                                (f^{<+}_{\bm k})^*  - (> \leftrightarrow <) \right).
\label{eq:zeroth_moment_normal_branch}
\end{equation}
To properly evaluate the first moment, we need to address the time-derivative in the second term of~\cref{eq:first_moment}. We work to the lowest order in gradients, whereby the time-derivative acts only on the shell-function triplets inside the self-energy functions~\cref{eq:Pi-in_4}. For these, we use the evolution equation~\cref{eq:eom_for_fs}, which implies $\partial_t f^{s+}_{\bm k} = 2i\omega_{\bm k}f^{s-+}_{\bm k}$ and $\partial_t (f^{s+}_{\bm k})^* = -2i\omega_{\bm k}f^{s+-}_{\bm k}$. In this way, we find
\begin{align}
    \langle C_{1\bm k} \rangle = \frac{\lambda^2}{4\omega_{\bm k}} \int {\rm dPS}_{\bm k}\; 
    \text{Re}\Big[ 
      &\omega_{{\bm k}_1}f^{>-+}_{{\bm k}_1} f^{>+}_{{\bm k}_2} 
                        (f^{<+}_{{\bm k}_3})^*(f^{<+}_{{\bm k}})^*
    \nonumber\\[-0.2ex] 
    + &\omega_{{\bm k}_2}f^{>+}_{{\bm k}_1} f^{>-+}_{{\bm k}_2} 
                        (f^{<+}_{{\bm k}_3})^*(f^{<+}_{{\bm k}})^*
    \nonumber\\[0.8ex] 
    - &\omega_{{\bm k}_3}f^{>+}_{{\bm k}_1} f^{>+}_{{\bm k}_2} 
                        f^{<+-}_{{\bm k}_3}(f^{<+}_{{\bm k}})^*
    \nonumber\\[0.3ex] 
    - &\omega_{\bm k}    f^{>+}_{{\bm k}_1} f^{>+}_{{\bm k}_2} 
                        (f^{<+}_{{\bm k}_3})^*f^{<++}_{{\bm k}} \Big] 
       - (>\leftrightarrow<).
    \label{eq:normal_C1}
\end{align}
Because the shell functions $f^{s,ab}_{\bm k}$ can be solved in terms of moments using~\cref{eq:fpmpm,eq:fpmmp} and~\cref{eq:useful_property_0}, the evolution equations for the moments are now closed in this case. 

%%%%%%%%%%%%%%%%%%%%%%%%%%%%%%%%%%%%%%%%%%%%%%%%%%%%%%%%%%%%%%%%%%%%%%%%%%%%%%%%%%%%%%%%%%%%%%%%%%%%%%%
\paragraph{Boltzmann limit.}
%%%%%%%%%%%%%%%%%%%%%%%%%%%%%%%%%%%%%%%%%%%%%%%%%%%%%%%%%%%%%%%%%%%%%%%%%%%%%%%%%%%%%%%%%%%%%%%%%%%%%%%

It is instructive to reduce our moment equations to the Boltzmann limit. To do this, one has to assume the spectral limit everywhere, and not only in the collision integrals as we have done so far. In the Boltzmann limit there is no quantum-coherence, {\em e.g.}~$f^{s\pm\mp}_{\bm k} \rightarrow 0$, whereby $\mathrm{Im}\langle C_{0\bm k}\rangle$ vanishes in this limit as well along with the first three lines in $\langle C_{1\bm k}\rangle$. Also, in the last term in $\langle C_{1\bm k}\rangle$, which solely arises from the first term in equation~\cref{eq:first_moment}, we must set $f^{s+}_{{\bm k}_i} \rightarrow f^{s++}_{{\bm k}_i}$. Then, differentiating~\cref{eq:fpmpm} for $f^{<++}_{\bm k}$ with respect to time and using equation~\cref{eq:rho2_eom} to lowest order in mass gradients, we find the standard Boltzmann equation:
\begin{equation}
\partial_t f^{<++}_{\bm k} \approx - \frac{\lambda^2}{4} \int {\rm dPS}_{\bm k}\; %_{1\{1\}} \; 
  \left( f^{>++}_{{\bm k}_1} f^{>++}_{{\bm k}_2} 
         f^{<++}_{{\bm k}_3} f^{<++}_{\bm k} 
         - (> \leftrightarrow <) \right).
\label{eq:boltzmann_limit}
\end{equation}
Note that equation~\cref{eq:boltzmann_limit} has right signs for the loss (first) and gain (second) terms. This is a good consistency check for the sign of the collision integrals in the full moment equations. The equation for negative frequency solutions $f^{<--}_{\bm k}$ reduces to~\cref{eq:boltzmann_limit}, as real scalar field particles are their own antiparticles.

%%%%%%%%%%%%%%%%%%%%%%%%%%%%%%%%%%%%%%%%%%%%%%%%%%%%%%%%%%%%%%%%%%%%%%%%%%%%%%%%%%%%%%%%%%%%%%%%%%%%%%%
\paragraph{Stable tachyonic branch.}
%%%%%%%%%%%%%%%%%%%%%%%%%%%%%%%%%%%%%%%%%%%%%%%%%%%%%%%%%%%%%%%%%%%%%%%%%%%%%%%%%%%%%%%%%%%%%%%%%%%%%%%

In the case $M^2_{\rm eff}<0$, additional kinematical channels stay open in~\cref{eq:Pi-in_2}, although the normal channels discussed above usually remain the dominant ones. The moments of the collision integral can still be written in an apparently simple form using our shorthand notations for the sums. The contribution from stable tachyons to the zeroth moment of the collision integral, in the case of a stable $\bm k$-mode, now reads: 
\begin{equation}
    \langle C_{0{\bm k}}(t)\rangle^{0,+}
= -\frac{i\lambda^2}{12\omega_{\bm k}} 
  \sum_{\{b_i\}}\,\int_{|M_{\rm eff}|}^\infty {\rm dPS}^{\bm k}_{1,\{b_i\}} \; 
                 {\rm Im} \left(f^{>b_1}_{{\bm k}_1} 
                                f^{>b_2}_{{\bm k}_2} 
                               (f^{<b_3}_{{\bm k}_3})^*
                                (f^{<+}_{\bm k})^*  - (> \leftrightarrow <) \right).
\label{eq:zeroth_moment_stable_tachyons}
\end{equation}
The limits in the integral, $|{\bm k}_i|\ge|M_{\rm eff}|$, apply to all three-momentum integrals in the phase space. In the first moment we use the equation of motion~\cref{eq:eom_for_fs} in the generic form to obtain a similarly restricted contribution from stable tachyons only:
\begin{align}
    \langle C_{1\bm k} \rangle^{0,+} 
    = \frac{\lambda^2}{12\omega_{\bm k}} 
    \sum_{\{b_i\}}\,\int_{|M_{\rm eff}|}^\infty {\rm dPS}^{\bm k}_{1,\{b_i\}} \; 
    \text{Re}\Big[ 
      &\omega^{b_1}_{{\bm k}_1}f^{>-b_1,b_1}_{{\bm k}_1} f^{>b_2}_{{\bm k}_2} 
                        (f^{<b_3}_{{\bm k}_3})^*(f^{<+}_{{\bm k}})^*
    \nonumber\\[-2ex]
    + &\omega^{b_2}_{{\bm k}_2}f^{>b_1}_{{\bm k}_1} f^{>-b_2,b_2}_{{\bm k}_2} 
                        (f^{<b_3}_{{\bm k}_3})^*(f^{<+}_{{\bm k}})^*
    \nonumber\\[1.2ex]
    - &\omega^{b_3}_{{\bm k}_3}f^{>b_1}_{{\bm k}_1} f^{>b_2}_{{\bm k}_2} 
                        f^{<b_3,-b_3}_{{\bm k}_3}(f^{<+}_{{\bm k}})^*
    \nonumber\\[0.7ex]
    - &\omega_{\bm k}    f^{>b_1}_{{\bm k}_1} f^{>b_2}_{{\bm k}_2} 
                        (f^{<b_3}_{{\bm k}_3})^*f^{<++}_{{\bm k}} \Big] 
       - (>\leftrightarrow<).
    \label{eq:stable_C1}
\end{align}
The indices in~\cref{eq:zeroth_moment_stable_tachyons,eq:stable_C1} indicate that the number of unstable internal momenta ${\bm k}_i$ is zero, and that also the moment ${\bm k}$ is stable (the sign of $\omega_{\bm k}^2$ is positive). Other possible channels will be found in section~\cref{subsec:unstable_tarchyon_collisions}. Even in this particular channel the compact notation conceals some complexity that must be dealt with in numerical work, as the new kinematical channels fall into three subsets, which can be evaluated separately. The explicit expressions are given in equations~\cref{eq:C0_stable_tachyon,eq:C1_stable_tachyon} in the appendix, where we also give the explicit collision integrals for the cases involving unstable tachyonic modes.

%%%%%%%%%%%%%%%%%%%%%%%%%%%%%%%%%%%%%%%%%%%%%%%%%%%%%%%%%%%%%%%%%%%%%%%%%%%%%%%%%%%%%%%%%%%%%%%%%%%%%%%
\paragraph{Coherence damping}
%%%%%%%%%%%%%%%%%%%%%%%%%%%%%%%%%%%%%%%%%%%%%%%%%%%%%%%%%%%%%%%%%%%%%%%%%%%%%%%%%%%%%%%%%%%%%%%%%%%%%%%

To finish this chapter, we take a closer look at the role of Im$\langle C_{0\bm k} \rangle$ in evolution equation~\cref{eq:rho0_eom} for a stable $M^2_{\text{eff}} > 0$. As stated before, the coherence shell solutions $f^{<\pm\mp}_{\bm k}$ vanish in the Boltzmann-limit. This condition can be expressed in terms of moments as $\rho_{2{\bm k}} = \omega_{\bm k}^2 \rho_{0{\bm k}}$ and $\partial_t\rho_{1{\bm k}}=0$. In the absence of collisions and assuming a slowly varying $\rho_{2\bm k}$, the equations of motion for $\rho_{0{\bm k}}$ and $\rho_{0\bm k}'$ describe harmonic oscillations around this state. There is no way to get rid of the quantum correlations arising from coherent particle production and the oscillations in $\rho_{0\bm k}$ continue indefinitely. Let us then consider the leading order term of $\langle C_{0\bm k} \rangle$ in the limit of a small deviation from the vacuum and slowly evolving moments, which together imply $|f^{<+}_{\bm k}| \ll 1$. After writing $\langle C_{0\bm k} \rangle$ in terms of the moments, the leading term is
\begin{equation}
    {\rm Im}\langle C_{0\bm k} \rangle = -\frac{\lambda^2}{4\omega_{\bm k}}\int {\rm dPS}_{\bm k}
    \left[\rho_{0{\bm k}_1}' \alpha_2 + \alpha_1 \rho_{0{\bm k}_2}' + \rho_{0{\bm k}_3}'\alpha_{\bm k} + \alpha_3 \rho_{0{\bm k}}'\right] \,,
\end{equation}
where we have used the shorthand $\alpha_i \equiv \omega_{{\bm k}_i} \rho_{0{\bm k}_i} + \rho_{1{\bm k}_i}$. The zeroth moment in the Minkowski vacuum, which is a good approximation for UV-modes in our setup \cite{Kainulainen_2023}, is $\rho_{0\bm k,\text{vac}} = 1/2\omega_{\bm k}$ and canonical commutation relations require $\rho_{1\bm k} = -1/2$. Thus, $\alpha_i > 0$ for a perturbation above the vacuum. $\text{Im}\langle C_{0\bm k}\rangle$ can then be recognized as a friction term, consistent with the phenomenological addition to the moment equations in \cite{Kainulainen:2021eki}. This term dampens the oscillation of $\rho_{0\bm k}$ around its coherence-free value and eventually destroys the coherence as the thermalization of the system proceeds.

The generally damping nature of the collision integrals can be seen by writing~\cref{eq:CI_moments_partially_integrated_2} as
\begin{equation}
    \langle C_{\alpha \bm k} \rangle  
    = - \frac{1}{2\omega_{\bm k}}\sum_{ab} 
        \left(\Gamma^\alpha_{{\bm k},\rm out}(\omega^a_{\bm k}) \hat f^{<ab}_{\bm k}  
      - \Pi^{<\alpha}_{{\bm k},\rm out}(\omega^a_{\bm k})\delta_{ab}\right),
    \label{eq:CI_moments_partially_integrated_3}
\end{equation}
where we used shorthand $A^\alpha_{{\bm k},\rm out} \equiv (\frac{i}{2}\partial_t + \delta_{ab}\omega^a_{\bm k})^\alpha A_{{\bm k},\rm out}$ and of course $\Gamma = i(\Pi^>-\Pi^<)/2$. Let us also note that in thermal limit $\Pi_{\bm k,\rm out}^{<,\alpha} = \Gamma^\alpha_{{\bm k},\rm out}f_{{\rm BE},{\bm k}}$. Now suppose that the dynamical evolution creates some non-thermal structures in $f^{<ab}_{\bm k}$. Even though $\Gamma^{\alpha}_{{\bm k},\rm out}$ and $\Pi^{<,\alpha}_{{\bm k},\rm out}$ are non-thermal, they are integrated quantities, and hence always smoother than the perturbed $f^{<ab}_{\bm k}$. This is particularly evident for the nonperturbative tachyonic particle production mechanism, which creates huge number of states at infrared frequencies. It is then clear that for a positive interaction rate~\cref{eq:CI_moments_partially_integrated_3} tends to damp perturbations both on diagonal and off-diagonal elements of $f^{ab}_{\bm k}$, and hence in all moments. 

%%%%%%%%%%%%%%%%%%%%%%%%%%%%%%%%%%%%%%%%%%%%%%%%%%%%%%%%%%%%%%%%%%%%%%%%%%%%%%%%%%%%%%%%%%%%%%%%%%%%%%%
%%%%%%%%%%%%%%%%%%%%%%%%%%%%%%%%%%%%%%%%%%%%%%%%%%%%%%%%%%%%%%%%%%%%%%%%%%%%%%%%%%%%%%%%%%%%%%%%%%%%%%%
%
\section{Stable branch with Landau damping}
\label{sec:Landau_damping}
%
%%%%%%%%%%%%%%%%%%%%%%%%%%%%%%%%%%%%%%%%%%%%%%%%%%%%%%%%%%%%%%%%%%%%%%%%%%%%%%%%%%%%%%%%%%%%%%%%%%%%%%%
%%%%%%%%%%%%%%%%%%%%%%%%%%%%%%%%%%%%%%%%%%%%%%%%%%%%%%%%%%%%%%%%%%%%%%%%%%%%%%%%%%%%%%%%%%%%%%%%%%%%%%%

Before moving to the case of collisions involving unstable tachyons, it is useful to study the stable branch including Landau damping. We do not expect that damping is necessarily relevant for the tachyonic dark matter production, but the topic is interesting on its own right and we get to introduce some structures that will help us better understand the unstable modes. 

As discussed above in section~\cref{sec:2PI}, Landau damping affects the pole- and the background solutions by giving them a finite width. This does not affect the form of the moment equations, but it does affect the approximation we make to get a closure in collision integrals. Taking a cue of our earlier result~\cref{eq:hartree_solution}, we argue that the appropriate Ansatz is:
\begin{align}
\nonumber \\[-15pt]
i \Delta^{s}_{\bm k}(u^0,v^0;t) &= \sum_{a,b} e^{-i\omega^a_{\bm k}(u^0-t)-\frac{1}{2}\Gamma_{\bm k}|u^0-t|} i\Delta^{s,ab}_{\bm k}(t,t) e^{-i\omega^b_{\bm k}(t - v^0)-\frac{1}{2}\Gamma_{\bm k}|t-v^0|}. 
\label{eq:hartree_solution_width}
\end{align}
\vskip-4pt
\noindent
This corresponds to generalizing the free time-evolution operators in~\cref{eq:hartree_solution} to their damped versions with the physically sensible boundary condition. It is easy to see that the damped time evolution operators are related to the spectral function:
\begin{align}
\sum_a a\int {\rm d}k_0 e^{-i(\omega_{\bm k}^a-k_0) r_0 - \frac{1}{2}\Gamma_{\bm k} |r_0| } 
&=\sum_a\frac{a\Gamma_{\bm k}}{(k_0-\omega_{\bm k}^a)^2+(\sfrac{1}{2}\Gamma_{\bm k})^2}
\nonumber \\
&\approx 4\omega_{\bm k}\frac{k_0\Gamma_{\bm k}}{(k_0^2-\omega_{\bm k}^2)^2+(\omega_{\bm k}\Gamma_{\bm k})^2}
= 4\omega_{\bm k} {\cal A}_{\bm k},
\label{eq:yetanorhereq}
\end{align}
where the limit $\Gamma_{\bm k} \ll \omega_{\bm k}$ was assumed. When $\Gamma_{\bm k} \to 0$, this becomes $4\omega_{\bm k}\pi\epsilon(k_0) \delta(k_0^2-\omega_{\bm k}^2)$. From~\cref{eq:yetanorhereq} we also see that the direct space expression for the damped spectral function is
\begin{equation}
{\cal A}_{\bm k}(u_0,v_0) = \frac{i}{2\omega_{\bm k}}\sin(\omega_{\bm k}(u_0-v_0)) e^{-\frac{1}{2}\Gamma_{\bm k} |u_0-v_0|}.
\label{eq:stable_spectral_function_with_width}
\end{equation}
This expression is antisymmetric in $u_0\leftrightarrow v_0$ and satisfies the sum-rule $2i\partial_{u_0}{\cal A}_{\bm k}(u_0,v_0)|_{u_0=v_0} = 1$. The Wigner transform of~\cref{eq:hartree_solution_width} can similarly be computed, with the result
\begin{align}
i\Delta^s_{\bm k}(k_0,t;\delta t) 
    = \sum_{ab} i\Delta^{ab}_{\bm k}(t,t)e^{2i\Delta\omega^{ab}_{\bm k}\delta t - \Gamma_{\bm k}|\delta t|} 
     \Big(\frac{\frac{1}{2}\Gamma_{\bm k} e^{2i(k_0-\bar\omega^{ab}_{\bm k})|\delta t|}}
                   {(k_0-\bar\omega^{ab}_{\bm k})(k_0-\bar\omega^{ab}_{\bm k} +\frac{i}{2}\Gamma_{\bm k})}
       + h.c. \Big),
\label{eq:full_damped_wigner_propagator}
\end{align}
where $\delta t \equiv t-\frac{1}{2}(u_0-v_0)$. The term in brackets in~\cref{eq:full_damped_wigner_propagator} can also be written as
\begin{equation}
\Big(...\Big) = \frac{\Gamma_{\bm k}}{(k_0-\bar\omega^{ab}_{\bm k})^2 + (\frac{1}{2}\Gamma_{\bm k})^2}
  \left( \cos (2(k_0-\bar\omega^{ab}_{\bm k})|\delta t|) + \Gamma_{\bm k}\frac{\sin (2(k_0-\bar\omega^{ab}_{\bm k})|\delta t|)}{2(k_0-\bar\omega^{ab}_{\bm k})} \right).
\label{eq:alternative_form}
\end{equation}
This form shows, that $i\Delta^s_{\bm k}(k_0,t;\delta t)$ collapses to eq.~\cref{eq:on_shell_wigner} in the spectral limit $\Gamma_{\bm k} \to 0$. Moreover,~\cref{eq:alternative_form} approaches the spectral form $2\pi\delta(k_0-\bar\omega^{ab}_{\bm k})$ also for a finite $\Gamma_{\bm k}$ in the long-time limit $(k_0-\bar\omega^{ab}_{\bm k})|\delta t| \gg 1$, where the sine term in~\cref{eq:alternative_form} becomes a delta-function and the oscillating cosine term becomes effectively zero. 

%%%%%%%%%%%%%%%%%%%%%%%%%%%%%%%%%%%%%%%%%%%%%%%%%%%%%%%%%%%%%%%%%%%%%%%%%%%%%%%%%%%%%%%%%%%%%%%%%%%%%%%
\paragraph{Moments including a finite width.}
%%%%%%%%%%%%%%%%%%%%%%%%%%%%%%%%%%%%%%%%%%%%%%%%%%%%%%%%%%%%%%%%%%%%%%%%%%%%%%%%%%%%%%%%%%%%%%%%%%%%%%%

Given a finite width, the convergence of the moment integrals~\cref{eq:moment_def_stable} is not obvious. Indeed, while all moments are well defined and finite for the spectral solution~\cref{eq:on_shell_wigner}, only the moments $n<2$ are well defined for~\cref{eq:full_damped_wigner_propagator}. To see this, we define moments with a limiting process $\delta t \to 0$, starting from~\cref{eq:full_damped_wigner_propagator} and noting that we need to keep a nonzero $\delta t$ only in the exponential in the integrand\footnote{We could equally well put $\delta t=0$ and start with a finite $u_0-v_0$ instead. The essential thing is to stay slightly away from $(u_0,v_0)=(t,t).$}:
\begin{align}
\rho_{n\bm k}(t) &= \lim\limits_{\delta t \to 0} \sum_{ab} i\Delta^{<ab}_{\bm k}(t,t) 
            \frac{1}{2}\int \frac{{\rm d}k_0}{2\pi}\Big(\frac{k_0^n \Gamma_{\bm k} e^{2i(k_0-\bar\omega^{ab}_{\bm k}|\delta t|}}
           {(k_0-\bar\omega^{ab}_{\bm k})(k_0-\bar\omega^{ab}_{\bm k} +\frac{i}{2}\Gamma_{\bm k})} + h.c. \Big).
\noindent \\
&\equiv \lim\limits_{\delta t \to 0} \sum_{ab} i\Delta^{<ab}_{\bm k}(t,t) I^{ab}_{n\bm k}(|\delta t|)
\label{eq:moment_wigner_propagator}   
\end{align}
As usual, we evaluate~\cref{eq:moment_wigner_propagator} with contour integration, breaking $I^{ab}_{n\bm k} = I^{ab}_{n\bm k,\rm res} + I^{ab}_{n\bm k,\rm arc}$. It is easy to see that the residues at $k_0=\omega^{ab}_{\bm k}$  give $I^{ab}_{n\bm k,\rm res} = (\bar\omega^{ab}_{\bm k})^n$ in our finite $\delta t$-regularization. This is the expected result, but the issue is with the convergence of the arc-integral. For $k_0 \equiv Re^{i\varphi}$ with $R\gg,\omega_{\bm k}, \Gamma_{\bm k}$, the latter can be written as  
\begin{align}
 I^{ab}_{n\bm k}(|\delta t|)_{\rm arc} 
&= \frac{1}{4\pi}\Gamma_{\bm k} R^{n-1} 
                  \sum_\pm e^{\mp 2i\bar\omega^{ab}_{\bm k}|\delta t|}\int_0^\pi {\rm d}\varphi 
                  e^{\pm i(n-1)\varphi \pm 2iR\cos(\varphi )|\delta t| \mp 2R\sin(\varphi )|\delta t|} 
 \nonumber \\
 & =\frac{1}{2\pi}\Gamma_{\bm k} R^{n-2} \cos(2R|\delta t|) \frac{\sin(2\bar\omega^{ab}_{\bm k}|\delta t|)}{|\delta t|} 
 %\xrightarrow[\bar\omega^{ab}_{\bm k}|\delta t| \ll 0]{}
 \to \frac{1}{\pi}\omega^{ab}_{\bm k}\Gamma_{\bm k} R^{n-2}\cos(2R|\delta t|).
\end{align}
where the limit $\bar\omega^{ab}_{\bm k}|\delta t| \ll 0$ was assumed in last step. The arc-integral then vanishes for $n=0,1$ and blows up for $n>2$. For $n=2$, it is finite and bound by $I^{ab}_{{n\bm k},\rm arc}<\frac{1}{\pi}\omega^{ab}_{\bm k}\Gamma_{\bm k} R^{n-2}$. However, since the arc-integral vanishes for any non-integer $n<2$ arbitrarily close to $n=2$, the second moment can still be sensibly defined as the sum of residues by this limiting procedure.

It still looks odd that moments are restricted to $n\le 2$ even for an infinitesimal $\Gamma_{\bm k}$, while all moments are well defined if one sets $\Gamma_{\bm k}\to 0$ first. This is related to the restrictive form of the local Ansatz~\cref{eq:hartree_solution_width}, leading to~\cref{eq:full_damped_wigner_propagator}. The problem can be ameliorated if one takes~\cref{eq:full_damped_wigner_propagator} as the starting point with the following generalization
\begin{align}
i\Delta^s_{\bm k}(k_0,t;\delta t) 
    &= \sum_{ab} e^{2i\Delta\omega^{ab}_{\bm k}\delta t - \Gamma_{\bm k}|\delta t|} 
     \Big(\frac{\frac{1}{2}\Gamma_{\bm k} e^{2i(k_0-\bar\omega^{ab}_{\bm k})|\delta t|}
     f^s_{\bm k}(k_0,t,\delta t)}
                   {(k_0-\bar\omega^{ab}_{\bm k})(k_0-\bar\omega^{ab}_{\bm k} +\frac{i}{2}\Gamma_{\bm k})}
       + h.c. \Big),
   \nonumber \\
   &\xrightarrow[\delta t \to 0]{} \sum_{ab} \frac{\Gamma_{\bm k} f^s_{\bm k}(k_0,t)}{(k_0-\bar\omega^{ab}_{\bm k})^2 + (\frac{1}{2}\Gamma_{\bm k})^2},
\label{eq:full_damped_wigner_propagator_extended}   
\end{align}
where we replaced the discrete coefficient functions $i\Delta^{ab}_{\bm k}(t,t)$ by a continuous function of frequency $f^s_{\bm k}(k_0,t)$. We recover the results of the previous approximation assuming that $f^s_{\bm k}(k_0,t)$ has no significant poles or cuts in complex plane. Indeed, if $f_{\bm k}^s(k_0,t)$ vanishes sufficiently fast for large $k_0$, then integrating either~\cref{eq:hartree_solution_width} or~\cref{eq:full_damped_wigner_propagator_extended} with any sufficiently smooth function of $k_0$ gives the same result with the identification $i\Delta^{s,ab}_{\bm k}(t,t) = f^{s}_{\bm k}(\bar\omega^{ab}_{\bm k}(t),t)$. In particular, all moments of~\cref{eq:full_damped_wigner_propagator_extended} can exist, not due to the convergence of the Breit-Wigner function, but due to that of $f^{s}_{\bm k}(k_0,t)$, for large $k_0$. This would naturally apply for the physical fluctuations in $f^{s}_{\bm k}(k_0,t)$, but for vacuum fluctuations, also necessarily contained in $f^{s}_{\bm k}(k_0,t)$, one is still restricted to $n\le 2$.
%

%%%%%%%%%%%%%%%%%%%%%%%%%%%%%%%%%%%%%%%%%%%%%%%%%%%%%%%%%%%%%%%%%%%%%%%%%%%%%%%%%%%%%%%%%%%%%%%%%%%%%%%
\paragraph{Collision integral with a finite width.}
%%%%%%%%%%%%%%%%%%%%%%%%%%%%%%%%%%%%%%%%%%%%%%%%%%%%%%%%%%%%%%%%%%%%%%%%%%%%%%%%%%%%%%%%%%%%%%%%%%%%%%%

To conclude, we show how the collision integral changes in our explicit example as a result of the finite width correction. Inserting the Ansatz~\cref{eq:hartree_solution_width} to~\cref{eq:Pi-in}, both explicitly and within the equation~\cref{eq:Pi-greater} for $i\Pi_{\bm k}^>(t,w_0)$, we see that the only change to the previous analysis occurs in the phase integral~\cref{eq:phase-integral-at-vertex}, which now becomes:
\begin{equation}
\int {\rm d}w_0 e^{-i(\sum_i \omega^{b_i}_{{\bm k}_i}-\omega^a_{\bm k})(w_0-t) - 2\Gamma_{\bm k} |w_0-t|} = \frac{4\Gamma_{\bm k}}{(\sum_i \omega^{b_i}_{{\bm k}_i}-\omega^a_{\bm k})^2+(2\Gamma_{\bm k})^2}.
\label{eq:breit-wigner-phase-space-integral}
\end{equation}
This translates to just replacing the frequency delta-function in~\cref{eq:Pi-greater} by the Breit-Wigner form~\cref{eq:breit-wigner-phase-space-integral}, (with the relabeling $b_3\to -b_3$). In a more general self-energy diagram, the energy conservation rule is similarly affected in each internal vertex. This loss of the sharp kinematic constraint would of course make collision integrals more challenging to evaluate in practice. This is a general result, not restricted to our simple scalar-field example. 

%%%%%%%%%%%%%%%%%%%%%%%%%%%%%%%%%%%%%%%%%%%%%%%%%%%%%%%%%%%%%%%%%%%%%%%%%%%%%%%%%%%%%%%%%%%%%%%%%%%%%%%
%%%%%%%%%%%%%%%%%%%%%%%%%%%%%%%%%%%%%%%%%%%%%%%%%%%%%%%%%%%%%%%%%%%%%%%%%%%%%%%%%%%%%%%%%%%%%%%%%%%%%%%
%
\section{Complete collision integrals in tachyonic branch}
\label{sec:ASA_approximation_unstable}
%
%%%%%%%%%%%%%%%%%%%%%%%%%%%%%%%%%%%%%%%%%%%%%%%%%%%%%%%%%%%%%%%%%%%%%%%%%%%%%%%%%%%%%%%%%%%%%%%%%%%%%%%
%%%%%%%%%%%%%%%%%%%%%%%%%%%%%%%%%%%%%%%%%%%%%%%%%%%%%%%%%%%%%%%%%%%%%%%%%%%%%%%%%%%%%%%%%%%%%%%%%%%%%%%

Evaluating tachyonic collision integrals is more intricate for interactions involving unstable modes with ${\bm k}^2 < M^2_\text{eff}$. These modes can be either growing or decaying in time, and the problem is that the growing modes do not have a particle interpretation in the constant mass limit. However, we need a reasonable modeling for them to guarantee the temporal continuity of the collision integrals in setups like~\cite{Fairbairn:2018bsw}, where low-momentum modes change from stable to unstable and back dynamically along the evolution. We start by studying general properties of the unstable modes, including finite temporal box regularization and finding the phase space structure of the converging unstable modes. In the end we use these results propose a simple prescription for computing collision integrals with unstable modes.

In this section, we shall assume that $\Gamma = 0$. We start by observing that in the unstable tachyonic region the spectral function is
\begin{equation}
    {\cal A}_{\bm k}(u_0,v_0) = - \frac{i}{2\tilde\omega_{\bm k}} \sinh(\tilde\omega_{\bm k}(u_0-v_0)),
\label{eq:unstable_spectral_function}
\end{equation}
where $\tilde\omega^2_{\bm k} = |M_{\rm eff}^2|-{\bm k}^2$. This result can be easily derived by a Laplace transformation of the collisionless equation for the spectral function: $(k_0^2 + \tilde\omega^2_{\bm k}){\cal A}_{\bm k}=0$ and imposing the anti-symmetry ${\cal A}_{\bm k}(u_0,v_0) =-{\cal A}_{\bm k}(v_0,u_0)$ and the spectral sum-rule $2i\partial_{u_0}{\cal A}_{\bm k}(u_0,v_0)|_{u_0=v_0} = 1$. Equation~\cref{eq:unstable_spectral_function} clearly corresponds to an analytic continuation $\omega_{\bm k}\rightarrow i\tilde\omega_{\bm k}$ of the stable branch solution~\cref{eq:stable_spectral_function_with_width}, (with $\Gamma_{\bm k} = 0$). This motivates us to consider the analytic continuation of~\cref{eq:hartree_solution} as the propagator to use for scattering unstable tachyonic modes:
\begin{align}
\nonumber \\[-15pt]
i \Delta^{s}_{\bm k}(u^0,v^0;t) &= \sum_{a,b} e^{\tilde\omega^a_{\bm k}(u^0-t)} i\Delta^{s,ab}_{\bm k}(t,t) e^{\tilde\omega^b_{\bm k}(t - v^0)}. 
\label{eq:hartree_solution_unstable}
\end{align}
\vskip-4pt
\noindent
where $\tilde\omega^a_{\bm k}\equiv a\tilde\omega_{\bm k}$. This is also just the solution of the free equation of motion~\cref{eq:hartree_equation} in the tachyonic regime.  However, the expression~\cref{eq:hartree_solution_unstable} is clearly divergent as $|u^0|,|v^0| \rightarrow \infty$. This is not surprising, as the solution~\cref{eq:hartree_solution_unstable} assumes a constant mass and dropping mass gradients implies that the mode remains tachyonic indefinitely, which leads to the observed infinite growth. However, in our physical setup this growth would be cut by the finite temporal width of tachyonic windows and because the positive backreaction terms in equation~\cref{eq:effective_mass} would eventually push $M_{\rm eff}^2$ to a positive value. 

%%%%%%%%%%%%%%%%%%%%%%%%%%%%%%%%%%%%%%%%%%%%%%%%%%%%%%%%%%%%%%%%%%%%%%%%%%%%%%%%%%%%%%%%%%%%%%%%%%%%%%%
\paragraph{Finite size temporal region.}
%%%%%%%%%%%%%%%%%%%%%%%%%%%%%%%%%%%%%%%%%%%%%%%%%%%%%%%%%%%%%%%%%%%%%%%%%%%%%%%%%%%%%%%%%%%%%%%%%%%%%%%

Motivated by the expected finite width of tachyonic windows, we study the solution~\cref{eq:hartree_solution_unstable} in a system confined to a finite time interval $[-T/2,T/2]$ around the time $t$. First, consider the stable case~\cref{eq:hartree_solution} with a real $\omega_{\bm k}$. The finite-box Wigner transform in this case gives
\begin{align}
    i\Delta^s_{\bm k}(k_0,x_0) 
    &= \sum_{ab} e^{2i\Delta\omega^{ab}_{\bm k}(t-x_0)}i\Delta^{ab}_{\bm k}(t,t) \int^{T/2}_{-T/2} dr_0 e^{i(k_0 - \bar\omega^{ab}_{\bm k})r_0} \nonumber\\
    &= \sum_{ab} e^{2i\Delta\omega^{ab}_{\bm k}(t-x_0)} i\Delta^{s,ab}_{\bm k}(t,t) \frac{2}{k_0 - \bar\omega^{ab}_{\bm k}}\sin\left( \frac{k_0 - \bar\omega_{\bm k}^{ab}}{2} T \right).
\label{eq:box_wightmann_stable}
\end{align}
This result clearly converges to~\cref{eq:on_shell_wigner} as $T\to\infty$. The corresponding solution for unstable modes with~\cref{eq:hartree_solution_unstable} reads 
\begin{align}
    i\Delta_{\bm k}(k_0,x_0) = \sum_a 
    \Biggl\{  &i\Delta^{aa}_{\bm k} \frac{2}{\tilde \omega^a_{\bm k} + i k_0} \sinh\left((\tilde \omega^a_{\bm k} + ik_0)\frac{T}{2} \right) 
   \nonumber \\ 
    + &i\Delta^{a,-a}_{\bm k} e^{-2\tilde \omega^a_{\bm k}(t-x_0)} 
    \frac{2}{k_0}\sin\left( \frac{k_0 T}{2} \right) \Biggr\},
\label{eq:box_wightmann_unstable}
\end{align}
The off-diagonal structure in~\cref{eq:box_wightmann_unstable} is similar to the stable case~\cref{eq:box_wightmann_stable} and well-defined in the limit $T \rightarrow \infty$. It is worth noting that only the off-diagonal part displays the unstable growth and decay in the mean time $x_0$, while the diagonal part diverges as a function of the relative time boundary $T$. Some sense can still be made of this solution by taking particular limits in the appropriate order:
\begin{equation}
    \frac{2}{\tilde \omega^a_{\bm k} + i k_0}
       \sinh\left( (\tilde \omega^a_{\bm k} + i  k_0)\frac{T}{2} \right) 
    \xrightarrow{\tilde\omega_{\bm k}\rightarrow 0} 
        \frac{2}{k_0}\sin\left( k_0 \frac{T}{2} \right) 
    \xrightarrow{T\rightarrow\infty} 2\pi \delta(k_0).
\label{eq:limiting_on_shell_coefficient}
\end{equation}
That is, if we can confine the dynamically relevant part of the collision integral within time $T$ such that $\tilde \omega_{\bm k} \ll T^{-1} \ll k_0$, a proper Ansatz for the collision integral would be to merge all shells together at $|\bm k| = |M_{\text{eff}}|$, as depicted in the right panel of Fig.~\cref{fig:dispersion_relations}. 

%=======================================================================================================
\begin{figure}[t]
   \centering
   \includegraphics[width=1.0\linewidth]{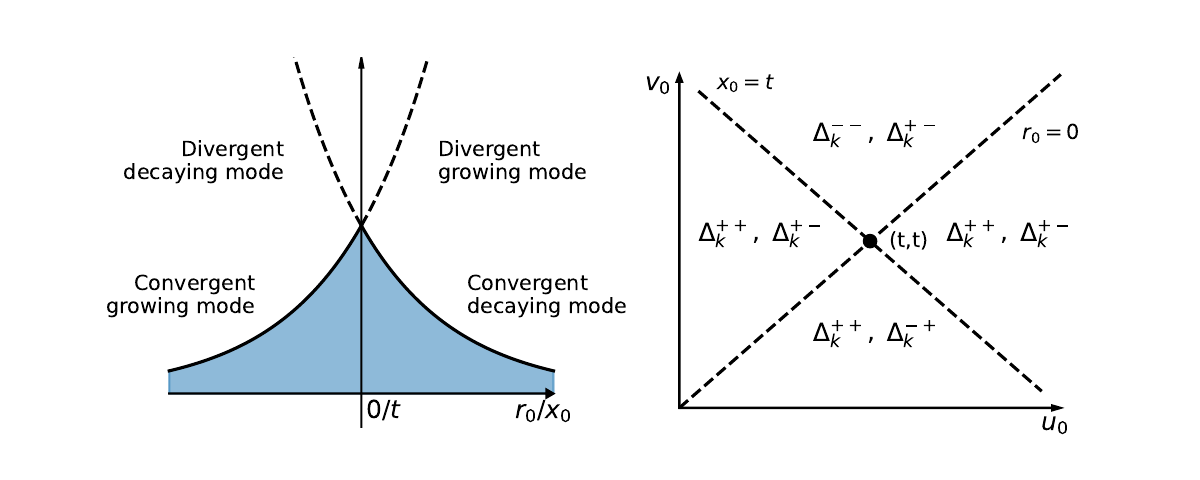} 
\caption{Left panel: The physical boundary conditions set for the correlation function in relative time (diagonal components) or mean time (off-diagonal components). Only the convergent parts of $\Delta^s$ give a finite contribution to the phase space integrals. Right panel: The components of the convergent part of the two-point functions each evolve in their respective quadrants of the $(u_0,v_0)$-plane.}
\label{fig:boundary_conditions_unstable}
\end{figure}
%=======================================================================================================

In practice, the relevant quantities to evaluate are the phase space integrals associated with the internal vertices in the self-energy diagrams. With our example self-energy term this means the integral~\cref{eq:phase-integral-at-vertex}, whose modification by Landau damping was considered in~\cref{eq:breit-wigner-phase-space-integral}. With the Ansatz~\cref{eq:hartree_solution_unstable} and the box-regularization this phase space integral becomes schematically:
\begin{equation}
\int_{-T}^T {\rm d}r_0 e^{-i\Omega r_0 - \tilde\Omega r_0} = \frac{2}{\tilde\Omega + i\Omega} \sinh\left((\tilde \Omega + i\Omega)\frac{T}{2} \right) 
\xrightarrow{} 2\pi \delta(\Omega),
\label{eq:unstable-phase-space-integral}
\end{equation}
where $\tilde\Omega$ denotes to the sum of frequencies of the unstable modes and $\Omega$ denotes the sum of stable modes involved in the vertex. The limit in~\cref{eq:unstable-phase-space-integral} was understood in the same sense as in~\cref{eq:limiting_on_shell_coefficient}, {\em e.g.}~assuming that $\tilde \Omega \ll T^{-1} \ll \Omega$. The physical assumptions going to this inequality are as follows: First, the unstable growth is assumed to be negligible within the correlation time and second, the wave-length of stable states involved in the process is much less than $T$, so that they can be treated as traveling waves.

%%%%%%%%%%%%%%%%%%%%%%%%%%%%%%%%%%%%%%%%%%%%%%%%%%%%%%%%%%%%%%%%%%%%%%%%%%%%%%%%%%%%%%%%%%%%%%%%%%%%%%%
\paragraph{Convergent unstable tachyon phase space.}
%%%%%%%%%%%%%%%%%%%%%%%%%%%%%%%%%%%%%%%%%%%%%%%%%%%%%%%%%%%%%%%%%%%%%%%%%%%%%%%%%%%%%%%%%%%%%%%%%%%%%%%

A complementary approach to study unstable tachyonic interactions is to restrict ourselves only to convergent modes. To this end, first note that the spectral function~\cref{eq:unstable_spectral_function} can be split into convergent and divergent parts,
\begin{equation}
    {\cal A}_{\bm k}(u_0,v_0) = \frac{i}{4\omega_{\bm k}} \text{sgn}(u_0-v_0) \sum_\pm \pm e^{\mp \omega_{\bm k} |u_0-v_0|}.
\label{eq:spectral_function_division}    
\end{equation}
It is interesting that the convergent and divergent parts each contain half of the sum rule and additional delta function terms, which cancel in the sum:
\begin{align}
    2i\partial_{u_0}\left[ \pm \frac{i}{4\omega_{\bm k}}\text{sgn}(u_0-v_0)e^{\mp \omega_{\bm k}|u_0-v_0|} \right]_{u_0 = v_0} = \frac12 \mp \frac{1}{\omega_{\bm k}}\delta(u_0-v_0).
\end{align}
The delta-functions arise from the discontinuity at $u_0=v_0$ in the restricted spectral functions~\cref{eq:spectral_function_division}. This suggests, that we should expect some spurious results also when considering interactions constrained to convergent solutions only. At any rate, the propagator~\cref{eq:hartree_solution_unstable} can be split into convergent and divergent parts as follows:
\begin{align}
    i\Delta^s_{\bm k}(r_0,x_0;t) &= \sum_{\pm,a} \left( 
    \theta(\pm ar_0) i\Delta^{saa}_{\bm k}(t,t) e^{\mp\tilde\omega_{\bm k}|r_0|} + \theta(\pm a(t-x_0)) i\Delta^{sa,-a}_{\bm k}(t,t) e^{\mp2\tilde\omega_{\bm k}|t-x_0|} \right) 
    \nonumber\\
    &\equiv i\Delta^s_{\bm k}(r_0,x_0;t)_{\text{conv}} + i\Delta^s_{\bm k}(r_0,x_0;t)_{\text{div}},
\end{align}
where converging solution corresponds to the upper and the diverging solution to the lower signs and we defined $r_0 = u_0-v_0$ and $x_0=(u_0+v_0)/2$. Both functions are continuous, but display cusps at $r_0=0$, as indicated in the left panel of figure~\cref{fig:boundary_conditions_unstable}. The convergent part of the correlation function can be Wigner-transformed, leading to a well defined phase space structure (at $x_0=t$):
\begin{equation}
    i\Delta^s_{\bm k}(k_0,x_0;t)_{\text{conv}} 
    = \frac{1}{\tilde\omega_{\bm k}}\left(\frac{f^{s++}_{\bm k}}{k_0-i\tilde\omega_{\bm k}}
          +\frac{f^{s--}_{\bm k}}{k_0+i\tilde\omega_{\bm k}}
          +\frac{2i\epsilon}{k_0^2+\epsilon^2}
          f^{s,c}_{\bm k} \right),
\label{eq:convergent_solution_phase_space}
\end{equation}
where $f^{s,c}_{\bm k} \equiv \sum_\pm f_{\bm k}^{s\pm\mp}$. The shell functions $f^{s,ab}_{\bm k}$ are still related to $i\Delta^{ab}_{\bm k}$ by~\cref{eq:fhat-functions}, with a complex frequency $\omega_{\bm k} = i\tilde\omega_{\bm k}$. This analytic continuation extends to all moments computed using~\cref{eq:convergent_solution_phase_space}, such that also the  relations~\cref{eq:rho0_cqpa,eq:rho1_cqpa,eq:rho2_cqpa,eq:rho0p_cqpa} and~\cref{eq:fpmpm,eq:fpmmp} hold for convergent unstable tachyons with a complex $\omega_{\bm k}= i\tilde\omega_{\bm k}$\footnote{Analogously to equation~\cref{eq:full_damped_wigner_propagator_extended}, the coefficient functions could be replaced with a single complex $k_0$-valued function, which is smooth enough so that only the explicit poles in~\cref{eq:convergent_solution_phase_space} contribute to frequency integrals.}. A phase-space structure similar to~\cref{eq:convergent_solution_phase_space} does {\em not} exist for growing modes. A Wigner transform for them does not converge with a constant mass, and so these states do not have an asymptotic particle interpretation. 

%%%%%%%%%%%%%%%%%%%%%%%%%%%%%%%%%%%%%%%%%%%%%%%%%%%%%%%%%%%%%%%%%%%%%%%%%%%
\paragraph{Collisions involving only convergent unstable tachyons.}
%%%%%%%%%%%%%%%%%%%%%%%%%%%%%%%%%%%%%%%%%%%%%%%%%%%%%%%%%%%%%%%%%%%%%%%%%%%

We next evaluate collision integrals constrained to convergent modes. This corresponds to a particular (unphysical) choice of boundary conditions, set separately for each quadrant around the point $(t,t)$ as shown in right panel of figure~\cref{fig:boundary_conditions_unstable}. Here the influence of the equal-time correlator $i\Delta^s_{\bm k}(t,t)$ decays in the future direction both in the mean and relative times, as do the influence on $i\Delta^s_{\bm k}(t,t)$ from events in the past direction. 

Consider the special case where all propagators within self-energies belong to the stable branch and only the external propagator with momentum ${\bm k}$ is unstable. In this case $t-x_0=(t-w_0)/2 = r_0/2$ and the converging unstable tachyon correlation function collapses to
\begin{equation}
i\Delta^<_{\bm k}(r_0;t) = \sum_{a,b} (i\Delta^{<ab}_{\bm k}) \theta(ar_0) e^{-\tilde\omega_{\bm k}|r_0|}.
\label{eq:collapsed_correlator}
\end{equation}
The relevant quantity is again the phase integral over the vertex time $w_0$. Given~\cref{eq:collapsed_correlator} it becomes:
\begin{equation}
\int_{-\infty}^\infty {\rm d}r_0 \theta(\pm r_0) e^{-i\Omega r_0 - \tilde\omega_{\bm k} |r_0|} 
= \frac{\tilde\omega_{\bm k}}{\Omega^2+\tilde\omega_{\bm k}^2}
  \mp \frac{i\Omega}{\Omega^2+\tilde\omega_{\bm k}^2} \xrightarrow[\tilde\omega_{\bm k}\to 0]{} \pi\delta(\Omega) \pm i{\textrm{PP}}\frac{1}{\Omega}, 
\label{eq:convergent-tachyon-space-integral}
\end{equation}
where $\Omega \equiv \sum_i \omega^{b_i}_{{\bm k}_i}$. In the limit $\tilde\omega_{\bm k}\to 0$ we thus find half of the delta-function discovered in the small $T$-limit using temporal box-regularization in~\cref{eq:unstable-phase-space-integral}, plus an additional principal value term. The latter must be spurious because it does not vanish in the $\tilde\omega_{\bm k}\to 0$ limit, where the box-regularization gives the correct result. Its appearance, related to the cusp in the convergent correlation function at $r_0=0$, is similar to the spurious $\delta$-functions found in the constrained spectral functions. 

%%%%%%%%%%%%%%%%%%%%%%%%%%%%%%%%%%%%%%%%%%%%%%%%%%%%%%%%%%%%%%%%%%%%%%%%%%%
\paragraph{The final prescription for unstable tachyons.}
%%%%%%%%%%%%%%%%%%%%%%%%%%%%%%%%%%%%%%%%%%%%%%%%%%%%%%%%%%%%%%%%%%%%%%%%%%%

The difference between the limiting forms of~\cref{eq:unstable-phase-space-integral} and ~\cref{eq:convergent-tachyon-space-integral} is clearly due to our ignoring the growing modes in deriving the latter. We do not have the means to evaluate this contribution directly, but the previous analysis indicates that to leading order it must add $\pi\delta(\Omega)\mp i{\rm PP}(1/\Omega)$ to~\cref{eq:convergent-tachyon-space-integral} in the $\tilde\omega_{\bm k}\to 0$ limit. This is reasonable, because convergent and divergent parts of the propagator obey the same boundary conditions at $r_0=0$, {\em e.g.}~they are parametrized by the same functions $i\Delta^{sab}_{\bm k}(t,t)$. We then conjecture that the best approximation we can make in the constant mass limit, is replacing the phase integrals by
\begin{equation}
\int_{-\infty}^\infty {\rm d}r_0 e^{-i\Omega r_0 - \tilde\Omega |r_0|}
= \frac{2\tilde\omega}{\Omega^2+\tilde\Omega^2}\quad \xrightarrow[\tilde\Omega\to 0]{} \quad 2\pi\delta(\Omega),
\label{eq:convergent-tachyon-space-integral_2}
\end{equation}
\vskip-0.2cm \noindent
where $\tilde\Omega$ and $\Omega$ are respectively the sums of the frequencies corresponding to unstable and stable modes entering a given internal vertex. Note the similarity of this result with~\cref{eq:breit-wigner-phase-space-integral} found in the stable branch including Landau damping, with $\tilde\Omega$ taking over the role of the width $\Gamma$.  In practice, we will always make the limiting approximation indicated in~\cref{eq:convergent-tachyon-space-integral_2} and replace the Breit-Wigner form with the $\delta(\Omega)$-function. This corresponds to setting the unstable tachyon frequencies to zero in the kinematic constraint. After all the trouble we went through, it is humbling to note that the limiting rule in~\cref{eq:convergent-tachyon-space-integral_2} is actually equivalent to using just the local limit
\begin{align}
\nonumber \\[-15pt]
i \Delta^{s}_{\bm k}(u^0,v^0;t) &\to \sum_{a,b} i\Delta^{s,ab}_{\bm k}(t,t), 
\label{eq:hartree_solution_unstable_2}
\end{align}
\vskip-4pt
\noindent
for the correlation function in the unstable tachyonic side. We warn, however, that the form~\cref{eq:hartree_solution_unstable_2} must be used in~\cref{eq:CI_moments_partially_integrated_2}, after the limit $r_0\to 0$ has already been taken.

%%%%%%%%%%%%%%%%%%%%%%%%%%%%%%%%%%%%%%%%%%%%%%%%%%%%%%%%%%%%%%%%%%%%%%%%%%%%%%%%%%%%%%%%%%%%%%%%%%%%%%%%%
%
\subsection{Collisions involving unstable tachyons}
\label{subsec:unstable_tarchyon_collisions}
%
%%%%%%%%%%%%%%%%%%%%%%%%%%%%%%%%%%%%%%%%%%%%%%%%%%%%%%%%%%%%%%%%%%%%%%%%%%%%%%%%%%%%%%%%%%%%%%%%%%%%%%%%%

The rule for unstable tachyons discussed above allows for a simple procedure to find generic collision integrals. These should be continuous across the tachyonic divide and this condition is indeed fulfilled automatically by~\cref{eq:hartree_solution_unstable_2} because $\omega_{\bm k}\to 0$ in~\cref{eq:hartree_solution}, when one approaches the tachyonic divide from the stable side.  In particular, we can then write the two lowest moments of the collision integral in our specific example in the following general forms:
\begin{align}
	\langle C_{0{\bm k}}\rangle  &= -\frac{i\lambda^2}{6} 
                         \sum_{a,\{b_i\}} \int \overbar{\rm dPS}^{\bm k}_{a,\{b_i\}} \; 
    {\rm Im}\left[ i\Delta^{>b_1}_{{\bm k}_1} i\Delta^{>b_2}_{{\bm k}_2} 
    i\Delta^{>b_3}_{{\bm k}_3} i\tilde\Delta^{<a}_{\bm k}\right]
\label{eq:unstable_C0}
\\
   \langle C_{1\bm k} \rangle &= -\frac{\lambda^2}{6} \sum_{a,\{b_i\}} \int \!
   \overbar{\rm dPS}^{\bm k}_{a,\{b_i\}} \; 
   \text{Re}\Big[\omega^a_{\bm k}
   \big[i\Delta^{>b_1}_{\bm k_1}i\Delta^{>b_2}_{\bm k_2}i\Delta^{>b_3}_{\bm k_3}\big] i\Delta^{<aa}_{\bm k} 
   \nonumber\\[-1.5ex]
  & \phantom{Hannatyttonenjaikuinen}
   + \Big(\frac{i}{2} \partial_t \big[i\Delta^{>b_1}_{\bm k_1}i\Delta^{>b_2}_{\bm k_2}i\Delta^{>b_3}_{\bm k_3}\big]\Big) 
   i\tilde \Delta^{<a}_{\bm k} \Big],
\label{eq:unstable_C1}
\end{align}
where $i\Delta^{s,b}_{\bm k}\equiv \sum_a i\Delta^{s,ab}_{\bm k}$, in analogy with equation~\cref{eq:summed_fs} and
\begin{equation}
\int \overbar{\rm dPS}^{{\bm k}}_{a,\{b_i\}} \equiv 
\int \prod_i \frac{{\rm d}^3 k_i}{(2\pi)^3}\, (2\pi)^4 \delta^{(3)} ({\bm k}-\textstyle{\sum_i}{\bm k}_i) \delta(\omega^a_{\bm k}-\textstyle{\sum_i}\omega^{b_i}_{{\bm k}_i}).
\label{eq:phase_space_2}
\end{equation}
Here, of course, the corresponding frequency is assumed to be zero for unstable tachyonic modes. A couple of comments are in order: In~\cref{eq:unstable_C0,eq:unstable_C1}, we used equation~\cref{eq:fhat-functions} in reverse order, so that the result is valid as such also for unstable tachyons for which $\omega_{\bm k}$ becomes complex. Moreover, we defined the summed correlator over the {\em second} index:
\begin{equation}
i\tilde \Delta^{<a}_{\bm k} \equiv \sum_b i\Delta^{<ab}_{\bm k}.
\end{equation} 
In the stable side we used $i\tilde \Delta^{<a}_{\bm k}=(i\Delta^{<a}_{\bm k})^*$, but on the unstable side this does not hold. The complex conjugation rule for $i\Delta^{ab}_{\bm k}$ changes from the $(i\Delta^{ab}_{\bm k})^*=i\Delta^{ba}_{\bm k}$ on the stable side (following from~\cref{eq:hartree_solution}) to $(i\Delta^{ab}_{\bm k})^*=i\Delta^{-b.-a}_{\bm k}$ on the unstable side (following from~\cref{eq:hartree_solution_unstable}). Due to the same reason, we also used the stable side relation $(i\Delta^{<b_3}_{{\bm k}_3})^* = i\Delta^{>-b_3}_{-{\bm k}_3}$ and relabeling $b_3\to-b_3$ and ${\bm k}_3\to-{\bm k}_3$ to bring $i\Delta^{<b_3}_{{\bm k}_3}$-term back to the more general form. In addition to being valid on both tachyonic regimes, equations~\cref{eq:unstable_C0,eq:unstable_C1} allow an easier identification of degeneracies in the kinematic constraints in what follows. 

Equations~\cref{eq:unstable_C0,eq:unstable_C1} contain a priori many combinations with different number of unstable tachyonic momenta. However, the kinematic constraints restrict the number of unstable legs to less than three. First, for three tachyonic momenta the phase-space frequency delta function vanishes identically, because it contains only a single non-zero frequency. Second, in the case with four unstable momenta the integrands in~\cref{eq:unstable_C0,eq:unstable_C1} vanish identically apart from the first term in~\cref{eq:unstable_C1}. This remaining term corresponds to a zero-measure integral when the tachyonic divide is approached from the stable side. From tachyonic side the term is parametrically consistent with zero within our scheme, as the continuity of collision integral requires. In the end, we can write the moments of collision integral schematically as follows
\begin{align}
\langle C_{\alpha\bm k} \rangle^+ &= \langle C_{\alpha\bm k} \rangle^{0,+} 
                              + \langle C_{\alpha\bm k} \rangle^{1,+} 
                              + \langle C_{\alpha\bm k} \rangle^{2,+}
\nonumber\\
\langle C_{\alpha\bm k} \rangle^- &= \langle C_{\alpha\bm k} \rangle^{0,-} 
                                   + \langle C_{\alpha\bm k} \rangle^{1,-} ~.
\label{eq:unstable_leg_expansions_for_C_alphas}
\end{align}
We recall that the $+(-)$ labels here correspond to $\bm{k}$ being  a stable (unstable) tachyonic mode, and the numerical indices on the right hand side denote the number of unstable internal momenta.
The terms $\langle C_{\alpha\bm k} \rangle^{0,+}$ were already given in~\cref{eq:zeroth_moment_stable_tachyons,eq:stable_C1}. They can be obtained from~\cref{eq:unstable_C0,eq:unstable_C1} using the derivative chain rule, constraining the momentum integrations to $|{\bm k}|>|M_{\rm eff}|$ and imposing the following replacement rules for all stable momentum states: %
\begin{equation}
{\rm Stable\!:} \qquad
i\Delta^{s,ab}_{\bm k} \to \frac{\hat f^{s,ab}_{\bm k}}{2\omega_{\bm k}}
\qquad {\rm and} \qquad
\frac{i}{2}\partial_t i\Delta^{s,c}_{\bm k} 
\to -\omega^{c}_{\bm k} f^{s,-c,c}_{\bm k}\frac{1}{2\omega_{\bm k}}.
\phantom{i}
\label{eq:replacement_rule_1}
\end{equation}
The first rule above is just~\cref{eq:fhat-functions} and the second follows from~\cref{eq:eom_for_fs}. These rules and the momentum constraint $|{\bm k}|>|M_{\rm eff}|$ should be used for all stable tachyonic momenta also in the other terms with unstable tachyonic modes in~\cref{eq:unstable_leg_expansions_for_C_alphas}. 

When a given momentum ${\bm k}$ is unstable, its momentum variable is restricted to range $|{\bm k}|<|M_{\rm eff}|$ and its frequency is eliminated from the frequency delta function in the phase space integral~\cref{eq:phase_space_2}. The latter operation allows one to perform the frequency sums over $b$, giving the simplified rules for unstable momenta:
\begin{equation}
{\rm Unstable\!:} \qquad
\sum_b i\Delta_{\bm k}^{>b} \to \rho_{0\bm k}
\qquad {\rm and} \qquad \sum_b \omega_{\bm k}^b i\Delta^{<bb}_{\bm k} \to 2\rho_{1{\bm k}} = -1.
\label{eq:replacement_rule_2}
\end{equation}
\vskip-0.3cm \noindent
The first rule is just the expression for the local correlator $\rho_{0\bm k}=i\Delta^<_{\bm k} = \sum_{ab}\Delta^{<ab}_{\bm k}$, which can be derived from any of the four relations~\cref{eq:hartree_solution,eq:hartree_solution_width,eq:hartree_solution_unstable,,eq:hartree_solution_unstable_2}. The second relation in~\cref{eq:replacement_rule_2} follows from~\cref{eq:rho1_cqpa} and is valid both for stable and unstable modes in our scheme. 

We give one example of collision integral moments involving unstable tachyons, and leave the rest of the cases to the appendix~\cref{sec:appendix_stable_branch_collision_integrals}. Indeed, consider the case where only the external momentum ${\bm k}$ is unstable. In this case the replacement rules~\cref{eq:replacement_rule_1,eq:replacement_rule_2} immediately give:
\begin{align}
	\langle C_{0{\bm k}}\rangle^{0,-}  &= -\frac{i\lambda^2}{6} 
                         \sum_{\{b_i\}} \int {\rm dPS}^{\bm k}_{0,\{b_i\}} \; 
    \rho_{0\bm k}\, {\rm Im}\big(f^{>b_1}_{{\bm k}_1} f^{>b_2}_{{\bm k}_2} 
    f^{>b_3}_{{\bm k}_3}\big)
\label{eq:unstable_C0_k_unstable}
\\
   \langle C_{1\bm k} \rangle^{0,-} &= \frac{\lambda^2}{6} \sum_{\{b_i\}} \int \!
   {\rm dPS}^{\bm k}_{0,\{b_i\}} \; 
   \text{Re}\Big(
   f^{>b_1}_{\bm k_1}f^{>b_2}_{\bm k_2}f^{>b_3}_{\bm k_3} 
   - \rho_{0\bm k}\frac{i}{2} \partial_t (f^{>b_1}_{\bm k_1}f^{>b_2}_{\bm k_2}f^{>b_3}_{\bm k_3}) \Big),
\label{eq:unstable_C1_k_unstable}
\end{align}
where we went back to using the phase-space element~\cref{eq:phase_space}: ${\rm dPS}^{\bm k}_{a,\{b_i\}}$, with $a=0$. Of course both of our explicit examples: $\langle C_{\alpha{\bm k}} \rangle^{0,\pm}$ still consist of several terms due to the sums over frequency variables. We leave both listing the remaining terms in~\cref{eq:unstable_leg_expansions_for_C_alphas}, as well as working out all kinematic channels within each term, to appendix~\cref{sec:appendix_stable_branch_collision_integrals}. \\

Finally, we wish to stress that our approximations involving unstable tachyonic modes do not in any way truncate the tachyonic growth, which is fully present in the dynamical evolution of the moment functions $\rho_{i\bm k}(t)$. What we {\em are} doing, is cutting away memory resulting from the exponential growth of unstable modes while $M_{\rm eff}^2<0$. Our rationale for doing so was based on the need to define local moments of the collision integral in closed form within the simple constant mass model. This approximation 
is motivated by the shortness of the tachyonic windows in our template setup~\cite{Fairbairn:2018bsw},
but in general its goodness depends on the system at hand. Of course, in systems where no tachyonic phases take place, such as a pure parametric resonance, the problem never arises. It would be interesting to study the issue further, for example using specific temporal mass profiles $M^2_{\rm eff}(t)$ for which exact solution can be found, but this goes beyond the scope of this work.

%%%%%%%%%%%%%%%%%%%%%%%%%%%%%%%%%%%%%%%%%%%%%%%%%%%%%%%%%%%%%%%%%%%%%%%%%%%%%%%%%%%%%%%%%%%%%%%%%%%%%%%
%%%%%%%%%%%%%%%%%%%%%%%%%%%%%%%%%%%%%%%%%%%%%%%%%%%%%%%%%%%%%%%%%%%%%%%%%%%%%%%%%%%%%%%%%%%%%%%%%%%%%%%
%
\section{Conclusions and discussion}
\label{sec:conclusions}
%
%%%%%%%%%%%%%%%%%%%%%%%%%%%%%%%%%%%%%%%%%%%%%%%%%%%%%%%%%%%%%%%%%%%%%%%%%%%%%%%%%%%%%%%%%%%%%%%%%%%%%%%
%%%%%%%%%%%%%%%%%%%%%%%%%%%%%%%%%%%%%%%%%%%%%%%%%%%%%%%%%%%%%%%%%%%%%%%%%%%%%%%%%%%%%%%%%%%%%%%%%%%%%%%

In this work, we used 2PI effective action methods to derive non-perturbative, renormalized quantum kinetic equations applicable for interacting scalar fields in a homogeneous and isotropic spacetime. The methods developed here are general, but as a concrete template for the scalar Lagrangian we have used the dark matter setup of~\cite{Fairbairn:2018bsw}, where scalar excitations are produced by tachyonic instabilities during reheating.

We started from the exact non-local Kadanoff-Baym equations and introduced four key approximations to reduce them to local kinetic equations, which still contain the relevant quantum effects. First, we showed that the equations describing the dynamics of the two-point functions can, to a good approximation, be decoupled from those governing the phase space structure of the system. Second, we performed next-to-leading order loop expansion in the self-energy components that enter collision terms, and leading order loop expansion in the Hermitian self-energy component, which defines the dispersive properties of the system. Third, following~\cite{Fidler:2011yq}, we moved to the Wigner space and derived equations of motion for $k_0$-moments of the two-point function. These moment functions are local in time, but the collision terms in their equations of motion still contain memory over the entire time evolution of the system. The fourth element in our approximation scheme was to reduce these memory-integrals to local limit by replacing the two-point functions {\em within them} by an Ansatz corresponding to lowest order expansion in gradients. We separately formulated the gradient expansion Ansatz for cases with positive and negative effective mass squared, and quantified its limitations in capturing deep infrared contributions of the collision integrals in the tachyonic regime. 

With this approach, we arrived at local equations of motion for the moments of the two-point function, renormalized  at leading order in the 2PI-expansion, and with next-to-leading order collision terms. These equations describe the evolution of scalar quantum systems with coherent structures in the presence of decohering, momentum exchanging interactions, and they can be solved numerically using methods developed for momentum-dependent Boltzmann equations. The computational resources required for the numerical implementation of our formalism are substantially less extensive than those required for direct lattice simulations of the 2PI equations. Moreover, the renormalization process in our approach is straightforward and, in contrast with the lattice methods, does not involve any additional numerical cost.
Our methods therefore provide a powerful tool for cosmological setups where out-of-equilibrium evolution of quantum fields has to be followed over a long period of time with a high phase-space resolution. The dark matter scenario of~\cite{Fairbairn:2018bsw} is a concrete example of such a setup, and we will return to the numerical application of our methods to this specific case in near future~\cite{KNV4}. Finally, our approach can also be straightforwardly generalized to more complicated systems, not only with more scalar fields but also including gauge fields and fermions~\cite{Jukkala:2019slc,Kainulainen:2024fdg}. 

A substantial part of this paper went to analyzing collision integrals involving unstable tachyonic modes, for which the the lowest order Ansatz for the 2-point-function is not always a consistent approximation. While the unstable tachyon contribution is not expected to be numerically large, their self-consistent treatment is important for numerical stability, {\em e.g.}~to ensure the continuity of collision terms, as infrared modes can cross from normal to tachyonic and back dynamically during the evolution of the system. After a careful analysis of converging unstable tachyonic modes and a general treatment using a temporal box-regularization, we were able to construct a simple  prescription for computing collision terms with all tachyonic modes. As a prelude to this problem, we also presented a concrete method for accurately evaluating collision integrals for excitations with a finite decay width. The main finding here was, that in the presence of a finite width the exact energy conservation at the interaction vertices is lost. In addition to direct numerical application of our methods and their extension to higher spin fields, it will be interesting to study the collisions with growing unstable tachyonic modes further, for example using some exactly solvable time dependent mass models.

%%%%%%%%%%%%%%%%%%%%%%%%%%%%%%%%%%%%%%%%%%%%%%%%%%%%%%%%%%%%%%%%%%%%%%%%%%%%%%%%%%%%%%%%%%%%%%%%%%%%%%%
%%%%%%%%%%%%%%%%%%%%%%%%%%%%%%%%%%%%%%%%%%%%%%%%%%%%%%%%%%%%%%%%%%%%%%%%%%%%%%%%%%%%%%%%%%%%%%%%%%%%%%%
%
\section*{Acknowledgements}
\label{sec:ack}
The work of OV was supported by a grant from the Magnus Ehrnrooth foundation.
%
%%%%%%%%%%%%%%%%%%%%%%%%%%%%%%%%%%%%%%%%%%%%%%%%%%%%%%%%%%%%%%%%%%%%%%%%%%%%%%%%%%%%%%%%%%%%%%%%%%%%%%%
%%%%%%%%%%%%%%%%%%%%%%%%%%%%%%%%%%%%%%%%%%%%%%%%%%%%%%%%%%%%%%%%%%%%%%%%%%%%%%%%%%%%%%%%%%%%%%%%%%%%%%%

%%%%%%%%%%%%%%%%%%%%%%%%%%%%%%%%%%%%%%%%%%%%%%%%%%%%%%%%%%%%%%%%%%%%%%%%%%%%%%%%%%%%%%%%%%%%%%%%%%%%%%%
%%%%%%%%%%%%%%%%%%%%%%%%%%%%%%%%%%%%%%%%%%%%%%%%%%%%%%%%%%%%%%%%%%%%%%%%%%%%%%%%%%%%%%%%%%%%%%%%%%%%%%%
%
\section*{Appendices}
\addcontentsline{toc}{section}{\protect\numberline{}Appendices}
\appendix
%
%%%%%%%%%%%%%%%%%%%%%%%%%%%%%%%%%%%%%%%%%%%%%%%%%%%%%%%%%%%%%%%%%%%%%%%%%%%%%%%

%%%%%%%%%%%%%%%%%%%%%%%%%%%%%%%%%%%%%%%%%%%%%%%%%%%%%%%%%%%%%%%%%%%%%%%%%%%%%%%
%
\section{Explicit collision integrals in the tachyonic region}
\label{sec:appendix_stable_branch_collision_integrals}
%
%%%%%%%%%%%%%%%%%%%%%%%%%%%%%%%%%%%%%%%%%%%%%%%%%%%%%%%%%%%%%%%%%%%%%%%%%%%%%%%%%%%%%%%%%%%%%%%%%%%%%%%
%%%%%%%%%%%%%%%%%%%%%%%%%%%%%%%%%%%%%%%%%%%%%%%%%%%%%%%%%%%%%%%%%%%%%%%%%%%%%%%%%%%%%%%%%%%%%%%%%%%%%%%

Here we list the explicit expressions for the collision integrals in the tachyonic regime with $M^2_{\rm eff}<0$. Evaluating the general expression  (\ref{eq:unstable_C0}) and  (\ref{eq:unstable_C1}) amounts to working out the kinematical conditions for each channel with a certain number of stable and unstable tachyonic modes, and then carrying out the remaining sums over frequency indices. 

The integration domain for $\langle C_{0\bm k} \rangle$ and $\langle C_{1\bm k} \rangle$ splits naturally into eight subdomains depending on if each of $\bm k_i$ is stable or not,
\begin{equation}
    \prod_{i=1}^3 \int d^3 k_i = \prod_{i=1}^3\Big(\int_0^{|M_{\text{eff}}|} d^3 k_i + \int_{|M_\text{eff}|}^\infty d^3 k_i\Big). 
\end{equation}
The eight domains can further be grouped into four distinct integrals by relabeling momenta. We use the compact notation introduced in the main text $\langle C_{\alpha\bm k} \rangle^{\pm, i}$, where $\pm$ refer to the sign of $\omega_q^2$, {\em e.g.}~$+$ ($-$) corresponds to stable (unstable) external mode ${\bf k}$ and $i$ counts the number of unstable internal modes ${\bf k}_i$. The full collision integrals are then simply
\begin{equation}
    \langle C_{\alpha\bm k}\rangle^{\pm}  = \sum_{i=0}^{n_\pm} \langle C_{\alpha\bm k}\rangle^{\pm,i} 
\end{equation}
where $n_+=2$ and $n_-=1$ as explained in the main text above~\cref{eq:unstable_leg_expansions_for_C_alphas}.
We also denote integrals over the phase space elements for stable tachyonic modes by 
\begin{equation}
\int {\rm d} \Pi^+_{i} \equiv \int_{|M_{\rm eff}|}^{\infty} \frac{{\rm d}^{3}k_i}{(2\pi)^3 2 
\omega_{{\bf k}_i}}~ 
\qquad {\rm and} \qquad
\int {\rm d} \Pi^-_{i} \equiv \int_0^{|M_{\rm eff}|} \frac{{\rm d}^{3}k_i}{(2\pi)^3}.
\label{eq:pshllkgg}
\end{equation}
Note that these shorthand integral measures have different dimensions, whereas ${\rm d} \Pi^+_{i}$ and ${\rm d} \Pi^-_{i}\rho_{0{\bm k}_i}$ do have the same dimension. The results for all the kinematically allowed channels for which the zeroth moment $\langle C_{0\bm k} \rangle^{\pm, i}$ is non-vanishing are given by 

\newcommand{\nsg}[1]{{\hspace{-2pt}{#1}\hspace{-0pt}}}
\newcommand{\nintd}{{\hspace{-2pt}\int\hspace{-3pt}{\rm d}}}
{\allowdisplaybreaks
\begin{align}
\nonumber     % ===============================================
\langle C_{0\bm k} \rangle^{0,+} = -\frac{i\lambda^2}{12\omega_{\bm k}} \;
\nintd \Pi_1^+ & {\rm d}\Pi_2^+ {\rm d}\Pi_3^+ \; 
(2\pi)^4 \delta^{(3)}({\bm k} -\Sigma_i{\bm k}_i) \times
\\[-0.5ex] \nonumber 
\times {\rm Im}\Big[
&\delta(\omega_{{\bm k}_1} \nsg{+} \omega_{{\bm k}_2}\nsg{+}
        \omega_{{\bm k}_3} \nsg{-} \omega_{{\bm k}})
   f^{>+}_{{\bm k}_1}  f^{>+}_{{\bm k}_2} 
   f^{>+}_{{\bm k}_3}  f^{>-}_{\bm k} 
\\ \nonumber
+ 3&\delta(\omega_{{\bm k}_1} \nsg{+} \omega_{{\bm k}_2} \nsg{-}
           \omega_{{\bm k}_3} \nsg{-} \omega_{{\bm k}})
    f^{>+}_{{\bm k}_1}      f^{>+}_{{\bm k}_2}
   f^{>-}_{{\bm k}_3} f^{>-}_{\bm k}
\\
+ 3&\delta(\omega_{{\bm k}_1} \nsg{+} \omega_{{\bm k}_2} \nsg{-}
           \omega_{{\bm k}_3} \nsg{+} \omega_{{\bm k}})
    f^{>+}_{{\bm k}_1}   f^{>+}_{{\bm k}_2}
   f^{>-}_{{\bm k}_3}f^{>+}_{\bm k}
-(> \leftrightarrow <) 
\Big], 
\label{eq:C0_stable_tachyon}
\\[1.2ex] \nonumber  % ===============================================
\langle C_{0\bm k} \rangle^{1,+} =-\frac{i\lambda^2}{4 \omega_{\bm k}} \;
\nintd\Pi_1^+ &{\rm d}\Pi_2^+ {\rm d}\Pi_3^- \;
(2\pi)^4 \delta^{(3)}({\bm k}-\Sigma_i{\bm k}_i) \; \times
\\[-0.5ex] \nonumber
\times \rho_{0{\bm k}_3} &{\rm Im}\Big[
\delta(\omega_{{\bm k}_1}\nsg{+} \omega_{{\bm k}_2}\nsg{-} 
\omega_{{\bm k}}) f^{>+}_{{\bm k}_1} f^{>+}_{{\bm k}_2} f^{>-}_{\bm k}
\\[-0.4ex]
&+2\delta(\omega_{{\bm k}_1}\nsg{-}\omega_{{\bm k}_2}\nsg{-}
\omega_{{\bm k}}) f^{>+}_{{\bm k}_1} f^{>-}_{{\bm k}_2}f^{>-}_{\bm k}
-(> \leftrightarrow <)
\Big],
\\[1.2ex]\nonumber % ===============================================
\langle C_{0\bm k} \rangle^{2,+} = -\frac{i\lambda^2}{4\omega_{\bm k}} \;
\nintd\Pi_1^+ &{\rm d}\Pi_2^- {\rm d} \Pi_3^- \;
(2\pi)^4 \delta^{(3)}({\bm k}-\Sigma_i{\bm k}_i)  \times
\\[-0.5ex]
\times \rho_{0{\bm k}_2 }&\rho_{0{\bm k}_3} \,
       \delta(\omega_{{\bm k}_1} \nsg{-}\omega_{{\bm k}}) \, 
       {\rm Im}\Big[f^{>+}_{{\bm k}_1} f^{>-}_{{\bm k}} 
       -(> \leftrightarrow <) \Big],
\\[1.2ex]\nonumber   % ===============================================
\langle C_{0\bm k} \rangle^{0,-} = -\frac{i\lambda^2}{2} \;
\nintd\Pi_1^+ &{\rm d}\Pi_2^+ {\rm d} \Pi_3^+ \;
(2\pi)^4 \delta^{(3)}({\bm k}-\Sigma_i{\bm k}_i)  \times
\\[-0.7ex]
\times \rho_{0{\bm k}} 
&\; \delta(\omega_{{\bm k}_1} \nsg{+} \omega_{{\bm k}_2} \nsg{-} \omega_{{\bm k}_3}) 
    {\rm Im}\Big[ f^{>+}_{{\bm k}_1} f^{>+}_{{\bm k}_2} f^{>-}_{{\bm k}_3})
    -(> \leftrightarrow <)\Big]~,
\\[1.2ex] \nonumber % ===============================================
\langle C_{0\bm k} \rangle^{1,-} =-\frac{i\lambda^2}{2} \;
\nintd\Pi_1^+ &{\rm d}\Pi_2^+ {\rm d} \Pi_3^- \;
(2\pi)^4 \delta^{(3)}({\bm k}-\Sigma_i{\bm k}_i) \,
\times
\\[-0.7ex]
\times \rho_{0{\bm k}} & \rho_{0{\bm k}_3} \delta(\omega_{{\bm k}_1} \nsg{-} \omega_{{\bm k}_2}) \; 
    {\rm Im}\Big[f^{>+}_{{\bm k}_1} f^{>-}_{{\bm k}_2} 
    - (> \leftrightarrow <) \Big].
\end{align} 
The channel where all four modes are unstable tachyons, $\langle C_{0\bm k}\rangle^{3,-}$ is kinematically allowed, but forward and backward terms cancel exactly in this case. The other two channels not listed above, $\langle C_{0\bm k}\rangle^{2,-}$ and $\langle C_{0\bm k}\rangle^{3,+}$, are kinematically forbidden.  In the positive mass squared region only the $\langle C_{0\bm k} \rangle^{0,+}$-term survives and within it, only the second delta function gives a surviving kinematic constraint. Then writing $f^{>-}_{\bm k}=(f^{<+}_{\bm k})^*$ and replacing the lower limit
in~\cref{eq:pshllkgg} by zero, one immediately recovers~\cref{eq:zeroth_moment_normal_branch}.

The corresponding results for the first moment $\langle C_{1\bm k} \rangle^{\pm, i}$ have a similar structure and one readily finds:
\begin{align} 
\nonumber
\langle C_{1\bm k} \rangle^{0,+} 
&= -\frac{\lambda^2}{12\omega_{\bm k}} \;
\nintd \Pi_1^+ {\rm d} \Pi_2^+ {\rm d} \Pi_3^+ \,
(2\pi)^4 \delta^{(3)}({\bm k}-\Sigma_i{\bm k}_i) \times
\\[0.2ex] \nonumber
&\times {\rm Re}\Big[
\delta(\omega_{{\bm k}_1} \nsg{+} \omega_{{\bm k}_2} \nsg{+}
       \omega_{{\bm k}_3} \nsg{-} \omega_{{\bm k}})
  \big(\omega_{{\bm k}} f_{{\bm k}_1}^{>+}f_{{\bm k}_2}^{>+}
                        f_{{\bm k}_3}^{>+}f_{{\bm k}}^{<++} 
   + \big[f_{{\bm k}_1}^{>+}f_{{\bm k}_2}^{>+} 
           f_{{\bm k}_3}^{>+} \big]' f_{{\bm k}}^{>-} \big)
\\[0.2ex] \nonumber
&\phantom{\times {\rm i}}
+ 3\delta(\omega_{{\bm k}_1} \nsg{+} \omega_{{\bm k}_2} \nsg{-}
          \omega_{{\bm k}_3} \nsg{-} \omega_{{\bm k}})
    \big(\omega_{{\bm k}}f_{{\bm k}_1}^{>+} f_{{\bm k}_2}^{>+}
                     f_{{\bm k}_3}^{>-} f_{{\bm k}}^{<++} 
    + \big[f_{{\bm k}_1}^{>+}f_{{\bm k}_2}^{>+}
            f_{{\bm k}_3}^{>-}\big]' f_{{\bm k}}^{>-}\big)
\\[0.5ex] \nonumber
&\phantom{\times {\rm i}}
+ 3\delta(\omega_{{\bm k}_1} \nsg{+} \omega_{{\bm k}_2} \nsg{-}
          \omega_{{\bm k}_3} \nsg{+} \omega_{{\bm k}})
  \big(\omega_{{\bm k}}f_{{\bm k}_1}^{>+}f_{{\bm k}_2}^{>+}
                       f_{{\bm k}_3}^{>-}f_{{\bm k}}^{>++} 
     + \big[f_{{\bm k}_1}^{>+}f_{{\bm k}_2}^{>+}
             f_{{\bm k}_3}^{>-}\big]' f_{{\bm k}}^{<+}\big)
\\
&\phantom{\times {\rm i}} - (> \leftrightarrow <) \Big],   
\\[1ex] \nonumber % ===============================================================
\langle C_{1\bm k} \rangle^{1,+} 
&= -\frac{\lambda^2}{4\omega_{\bm k}} \;
\nintd \Pi^+_1 {\rm d} \Pi^+_2 {\rm d} \Pi^-_3 \,
(2\pi)^4 \delta^{(3)}({\bm k}-\Sigma_i{\bm k}_i) \times
\\ \nonumber
&\times {\rm Re}\Big[
\delta(\omega_{{\bm k}_1} \nsg{+} \omega_{{\bm k}_2} \nsg{-}\omega_{{\bm k}})
\big(\omega_{\bm k} \rho_{0\bm k_3} f^{>+}_{\bm k_1} 
     f^{>+}_{\bm k_2} f^{<++}_{\bm k} 
     + [\rho_{0\bm k_3} f^{>+}_{\bm k_1} f^{>+}_{\bm k_2}]' \, f^{>-}_{\bm k} \big)
\\ \nonumber
&\phantom{\times {\rm i}}  
+2\delta(\omega_{{\bm k}_1}\nsg{-}\omega_{{\bm k}_2}
                           \nsg{-}\omega_{{\bm k}})
   \big( \omega_{\bm k} \rho_{0\bm k_3} f^{>+}_{\bm k_1} 
                      f^{>-}_{\bm k_2}  f^{<++}_{\bm k} 
  +  [\rho_{0\bm k_3} f^{>+}_{\bm k_1} f^{>-}_{\bm k_2}]' \, f^{>-}_{\bm k}\big) \\
&\phantom{\times {\rm i}} - (> \leftrightarrow <) \Big],
\\[1.2ex] \nonumber  % ===============================================================
\langle C_{1\bm k} \rangle^{2,+} 
&= -\frac{\lambda^2}{4\omega_{\bm k}} \;
\nintd \Pi_1^+ {\rm d} \Pi_2^- {\rm d} \Pi_3^- \,
(2\pi)^4 \delta^{(3)}({\bm k}-\Sigma_i{\bm k}_i) \;
\delta(\omega_{{\bm k}_1}\nsg{-}\omega_{{\bm k}}) \times
\\ 
&\times 
  \Big( \omega_{\bm k} \rho_{0\bm k_2}\rho_{0\bm k_3} 
      {\rm Re}\Big[ f^{>+}_{\bm k_1} f^{<++}_{\bm k}\Big] 
  -\frac{1}{2} \partial_t(\rho_{0\bm k_2}\rho_{0\bm k_3})\,
    {\rm Im}\Big[f^{>+}_{\bm k_1} f^{>-}_{\bm k} \Big]  - (> \leftrightarrow <) \Big),
\\[1.8ex]\nonumber  % ===============================================================
\langle C_{1\bm k} \rangle^{0,-} &= \frac{i\lambda^2}{2}
\nintd \Pi^+_1 {\rm d} \Pi^+_2{\rm d} \Pi^+_3 \,
(2\pi)^4 \delta^{(3)}({\bm k}-\Sigma_i{\bm k}_i) \,
\delta(\omega_{{\bm k}_1}\nsg{+} \omega_{{\bm k}_2}\nsg{-} \omega_{{\bm k}_3}) 
\times
\\[-0.5ex]
&\times\; 
       {\rm Re}\Big[ \Big(f^{>+}_{{\bm k}_1}f^{>+}_{{\bm k}_2} 
                          f^{>-}_{{\bm k}_3} + (> \leftrightarrow <) \Big)
       - \rho_{0\bm k} \Big( [f^{>+}_{{\bm k}_1}f^{>+}_{{\bm k}_2} 
                                                f^{>-}_{{\bm k}_3}]'
-(> \leftrightarrow <) \Big)\Big],
\\[1.8ex] \nonumber   % ===============================================================
\langle C_{1\bm k} \rangle^{1,-} &= \frac{i\lambda^2}{2}
\nintd \Pi^+_1 {\rm d} \Pi^+_2 {\rm d} \Pi^-_3 \,
(2\pi)^4 \delta^{(3)}({\bm k}-\Sigma_i{\bm k}_i) \,
\delta(\omega_{{\bm k}_1}\nsg{-} \omega_{{\bm k}_2}) 
\times
\\[-0.5ex]
&\times\; 
       {\rm Re}\Big[ \Big( \rho_{{\bm k}_3}f^{>+}_{{\bm k}_1}
       f^{>+}_{{\bm k}_2} + (> \leftrightarrow <) \Big)
       - \rho_{0\bm k} \Big( [\rho_{{\bm k}_3}f^{>+}_{{\bm k}_1}f^{>+}_{{\bm k}_2}]'
-(> \leftrightarrow <) \Big)\Big].
\label{eq:C1_stable_tachyon}
\end{align}
\vskip8pt\noindent
Here we defined a shorthand notation $[abc]'\equiv \frac{i}{2}\partial_t[abc]$, where the action of individual derivatives on shell functions is given by~\cref{eq:replacement_rule_1}. Explicitly:
\begin{align}
\big[f_{{\bm k}_1}^{>+}f_{{\bm k}_2}^{>+}f_{{\bm k}_3}^{>\pm}\big]' &= 
- \omega_{\bm k_1}f_{{\bm k}_1}^{>-+}f_{{\bm k}_2}^{>+}f_{{\bm k}_3}^{>\pm}
- \omega_{\bm k_2}f_{{\bm k}_1}^{>+}f_{{\bm k}_2}^{>-+}f_{{\bm k}_3}^{>\pm}
\mp \omega_{\bm k_3}f_{{\bm k}_1}^{>+}f_{{\bm k}_2}^{>+}f_{{\bm k}_3}^{>\mp\pm}
\label{eq:ruleA1}
\\
[\rho_{{\bm k}_3}f^{>+}_{{\bm k}_1}f^{>\pm}_{{\bm k}_2}]' &= 
\frac{i}{2}(\partial_t\rho_{{\bm k}_3}) f_{{\bm k}_1}^{>+} f_{{\bm k}_2}^{>\pm}
- \omega_{\bm k_1} \rho_{{\bm k}_3} f_{{\bm k}_1}^{>-+} f_{{\bm k}_2}^{>\pm} 
\mp \omega_{\bm k_2} \rho_{{\bm k}_3}f_{{\bm k}_1}^{>+} f_{{\bm k}_2}^{>\mp\pm}.
\label{eq:ruleA2}
\end{align}
Just as with $\langle C_{0{\bm k}} \rangle$, only the $\langle C_{1{\bm k}} \rangle^{0,+}$-term, and in it only its second delta-function contribution survives in the positive mass-squared case. Using~\cref{eq:ruleA1} along with $f^{>-}_{\bm k}=(f^{<+}_{\bm k})^*$ one then readily recovers~\cref{eq:normal_C1}.

In all expressions in this appendix, the unstable tachyon contributions are already written in terms of the moment functions. On the stable side one can use the relations~\cref{eq:fpmpm,eq:fpmmp} to write the shell functions in terms of moments. Indeed,~\cref{eq:fpmpm,eq:fpmmp} need only be supplemented by the relation $f^{>++}_{\bm k} = 1 + f^{<++}_{\bm k}$. In addition it is useful to write explicit expressions for the summed shell functions:
\begin{equation}
f^{<\pm}_{\bm k} = \omega_{\bm k}\rho_{0\bm k} \pm \Big(\rho_{1{\bm k}} - \frac{i}{2}\partial_t\rho_{0\bm k}\Big).
\label{eq:summed_fs_in_term_of_moments}
\end{equation}
Here of course $\rho_{1{\bm k}}=-1/2$. The summed $f^{>\mp}_{\bm k}$-functions follow from~\cref{eq:summed_fs_in_term_of_moments} using the relation
$f^{>\mp}_{\bm k}=(f^{<\pm}_{\bm k})^*$.

Finally, we point out that in resonant particle production the shell functions can become very large in the infrared region. At the same time the leading contributions cancel exactly between the forward and backward terms in all $\langle C_{\alpha{\bm k}}\rangle^{i,\pm}$. In order to achieve a good numerical accuracy, it is in practice useful to take care of this cancellation analytically. We do not show this explicitly here however, because it is a formally trivial task and it leads to somewhat more lengthy formulae than our compact expressions above.

%%%%%%%%%%%%%%%%%%%%%%%%%%%%%%%%%%%%%%%%%%%%%%%%%%%%%%%%%%%%%%%%%%%%%%%%%%%%%%%%%%%%%%%%%%%%%%%%%%%%%%%
%%%%%%%%%%%%%%%%%%%%%%%%%%%%%%%%%%%%%%%%%%%%%%%%%%%%%%%%%%%%%%%%%%%%%%%%%%%%%%%%%%%%%%%%%%%%%%%%%%%%%%%
%
\bibliography{maindesc.bib}
%
%%%%%%%%%%%%%%%%%%%%%%%%%%%%%%%%%%%%%%%%%%%%%%%%%%%%%%%%%%%%%%%%%%%%%%%%%%%%%%%%%%%%%%%%%%%%%%%%%%%%%%%
%%%%%%%%%%%%%%%%%%%%%%%%%%%%%%%%%%%%%%%%%%%%%%%%%%%%%%%%%%%%%%%%%%%%%%%%%%%%%%%%%%%%%%%%%%%%%%%%%%%%%%%

%%%%%%%%%%%%%%%%%%%%%%%%%%%%%%%%%%%%%%%%%%%%%%%%%%%%%%%%%%%%%%%%%%%%%%%%%%%%%%%%%%%%%%%%%%%%%%%%%%%%%%%
%%%%%%%%%%%%%%%%%%%%%%%%%%%%%%%%%%%%%%%%%%%%%%%%%%%%%%%%%%%%%%%%%%%%%%%%%%%%%%%%%%%%%%%%%%%%%%%%%%%%%%%
%
\end{document}